\documentclass[a4paper,11pt]{article}
\pdfoutput=1 
\usepackage{jheppub}   
\usepackage{graphicx}
\usepackage{float}
\usepackage{tensor}
\usepackage[toc,page]{appendix}
\usepackage[svgnames,table, usenames,dvipsnames]{xcolor}
\usepackage{multirow}
\usepackage{booktabs}
\usepackage{tabularx}

\newcommand{\be}{\begin{equation}}
\newcommand{\ee}{\end{equation}}
\newcommand{\ba}{\begin{eqnarray}}
\newcommand{\ea}{\end{eqnarray}}
\newcommand{\nn}{\nonumber}
\newcommand{\rt}{\tilde{r}_+}

\begin{document}

\title{Black holes in Einsteinian cubic gravity}

\author[a]{Robie A. Hennigar,}
\author[a]{and Robert B. Mann}

\affiliation[a]{Department of Physics and Astronomy, University of Waterloo, \\
Waterloo, Ontario, Canada, N2L 3G1}
\emailAdd{rhenniga@uwaterloo.ca}
\emailAdd{rbmann@uwaterloo.ca}

\abstract{
Using numerical and perturbative methods, we construct the first examples of black hole solutions in Einsteinian cubic gravity and study their thermodynamics.  Focusing first on four dimensional solutions, we show that these black holes have a novel equation of state in which the pressure is a quadratic function of the temperature.  Despite this, they undergo a first order phase transition with associated van der Waals behaviour.  We then construct perturbative solutions for general $D \ge 5$ and study the properties of these solutions for $D=5$ and $D=6$ in particular.   We note that for $D >4$ the solutions are described by two independent metric functions. We find novel examples of super-entropic behaviour over a large portion of the parameter space.  We analyse the specific heat, determining that the black holes are thermodynamically stable over large regions of parameter space.
}


\maketitle


\section{Introduction}

In recent years there has been considerable interest in the subject of higher curvature gravity, much of which has been motivated through attempts to provide a quantum description of the gravitational field.  For example, it was known as early as 1977 that supplementing the Einstein-Hilbert action with higher curvature interactions can lead to a renormalizable theory of gravity~\cite{Stelle:1976gc}.  It is generally expected that such corrections should appear in the low energy limit of the ultraviolet completion of gravity.  In the case of string theory this manifests through the appearance  of the Gauss-Bonnet term in corrections to the low energy effective action~\cite{Zwiebach:1985uq}.  More recently, higher curvature gravity has been of interest in holography, where corrections to the Einstein-Hilbert action allow for the study of a broader universality class of CFTs~\cite{Myers:2010ru, Myers:2010jv, Hofman:2009ug, Myers:2010tj, Brigante:2007nu}.

However, for all of the interest in higher curvature gravity, relatively few models have actually been explored. Generically, these theories are difficult to study due to higher derivative equations of motion and are often plagued with pathological properties.  For example, it is often the case that the linearized equations of motion describing the propagation of gravitons will reveal that these metric perturbations describe negative kinetic energy excitations, i.e. ghosts, signalling a breakdown of unitarity in the quantum theory~\cite{Boulware:1985wk}.  For these reasons, the primary focus of attention has been on Lovelock gravity~\cite{Lovelock:1971yv}, with some attention also devoted to quasi-topological gravity~\cite{Myers:2010ru, Myers:2010jv, Oliva:2010eb} and certain $f(\text{Lovelock})$ models~\cite{Bueno:2016dol} where these issues can be controlled.  

In addition to these models, some effort has also been devoted to constructing theories of gravity that are explicitly free of ghosts~\cite{Sisman:2011gz, Gullu:2014gza, Gullu:2015cha}.  Recently a new model of cubic curvature gravity has been presented which, when supplemented with quadratic and cubic Lovelock terms, is the unique cubic model of gravity that shares its graviton spectrum with Einstein gravity~\cite{Bueno:2016xff}  and has dimension-independent coupling constants.  The Lagrangian density of this theory---appropriately called \textit{Einsteinian cubic gravity}---is given by,
\be 
{\cal L} = \frac{1}{2 \kappa} \left[-2 \Lambda + R \right] + \beta_1 \chi_{4} + \kappa \left[\beta_2 \chi_{6} + \lambda {\cal P} \right]
\ee
where
\be 
{\cal P} = 12 \tensor{R}{_\mu ^\rho _\nu ^\sigma} \tensor{R}{_\rho ^\gamma _\sigma ^\delta}\tensor{R}{_\gamma ^\mu _\delta ^\nu} + R_{\mu\nu}^{\rho\sigma} R_{\rho\sigma}^{\gamma\delta}R_{\gamma\delta}^{\mu\nu} - 12 R_{\mu\nu\rho\lambda}R^{\mu \rho}R^{\nu\sigma} + 8 R_\mu^\nu R_\nu ^\rho R_\rho ^\mu
\ee
and $\chi_{4}$ and $\chi_{6}$ are the four- and six-dimensional Euler densities, respectively, and correspond to the usual Lovelock terms.  Interestingly, the new ${\cal P}$ contribution which is present in this model is neither trivial nor topological in four dimensions allowing for the study of the effects of cubic curvature terms in four dimensions.    In this paper we construct vacuum topological black hole solutions of this theory and study their thermodynamics, concentrating on the case where $\beta_1=\beta_2=0$ to study the effects of the new term alone.  The field equations of the theory are complicated fourth order differential equations and we have not been able to solve them analytically, but have resorted to perturbative methods to obtain solutions.  Notwithstanding the instabilities that higher order field equations could lead to (a problem which remains for future work), the fact that this is the unique theory with linearized equations matching Einstein gravity make it worthy of  study, since at the very least it should provide a useful holographic toy model.

Our thermodynamic analysis is performed within the context of \textit{black hole chemistry}~\cite{Kubiznak:2014zwa}.  In this framework the cosmological constant is promoted to a thermodynamic parameter~\cite{Creighton:1995au, Caldarelli:1999xj, Dolan:2010ha, Dolan:2011xt} in the first law of black hole mechanics, a result which is supported by geometric arguments~\cite{Kastor:2009wy, Kastor:2010gq}.  The thermodynamic interpretation of the cosmological constant is that of a pressure, and its conjugate quantity is termed the thermodynamic volume.  Studies employing this formalism have shown thermodynamic phenomena for black holes analogous to that observed in everyday systems, for example van der Waals behaviour~\cite{Kubiznak:2012wp}, triple points~\cite{Altamirano:2013uqa}, (multiple) reentrant phase transitions~\cite{Altamirano:2013ane,Frassino:2014pha}, isolated critical points~\cite{Dolan:2014vba, Hennigar:2015esa}, and most recently a superfluid phase transition~\cite{Hennigar:2016xwd}.  Similar results have been found in a large number of subsequent investigations; for example see 
\cite{Wei:2012ui, Cai:2013qga, Xu:2013zea, Mo:2014qsa, Wei:2014hba, Mo:2014mba, Zou:2014mha, Belhaj:2014tga, Xu:2014tja, Frassino:2014pha, Dolan:2014vba, Belhaj:2014eha, Sherkatghanad:2014hda, Hendi:2015cka, Hendi:2015oqa, Hennigar:2015esa, Hendi:2015psa, Nie:2015zia, Hendi:2015pda, Hendi:2015soe, Johnson:2015ekr, Hennigar:2015mco, Hendi:2016njy, Zeng:2016aly} for investigations focusing on higher curvature gravity, \cite{Cvetic:2010jb, Hennigar:2014cfa, Hennigar:2015cja} for entropy inequalities, \cite{Johnson:2014yja, Karch:2015rpa, Caceres:2015vsa, Dolan:2016jjc, Caceres:2016xjz} for extensions to gauge/gravity duality, and \cite{Kubiznak:2016qmn} for a general review. 

Our paper is organized as follows.  In the next section we study in detail black hole solutions to the four dimensional theory considering the solution both asymptotically and near a black hole horizon.  We then study the thermodynamics of the four dimensional black holes, finding that their equation of state is
a quadratic function of temperature, in contrast to all other black holes with spherical symmetry. We find nevertheless that they still exhibit  van der Waals behaviour and a reentrant phase transition.
 We then move on to consider solutions in higher, arbitrary dimensions.  We study the higher dimensional black hole solutions, finding that for $D>4$ two independent metric functions, $f(r)$ and $N(r)$, are required to satisfy the field equations.  We emphasize that, while considering the solution as a small coupling limit is possible, it is not valid for black holes of arbitrary horizon radius and we clarify when such a limit is sensible.  We finish by investigating the higher dimensional thermodynamics, focusing on entropy, the reverse isoperimetric inequality, and critical behaviour, providing details for five and six dimensions in particular.   Here we find that the higher dimensional black holes are thermodynamically stable over a large portion of parameter space and can be \textit{super-entropic}~\cite{Hennigar:2014cfa}.  The latter is the first such example for higher curvature black holes asymptotic to AdS space.  However, we find no interesting critical behaviour for $D=5$ and $D=6$ perturbatively in the coupling.

\section{Solutions in four dimensions}

In four dimensions the action of pure Einsteinian cubic gravity takes the form,
\be 
{\cal I} = \int d^4x \sqrt{-g}\left(\frac{1}{2\kappa} \left[-2 \Lambda + R \right] + \kappa \lambda {\cal P} \right) \, .
\ee

The vacuum Einstein equations which follow from the action can be conveniently written in the form,
\be 
G_{a b} = P_{a cde}R_{b}{}^{cde} - \frac{1}{2} g_{ab} {\cal L} - 2 \nabla^c \nabla^d P_{acdb} = 0
\ee
where
\begin{align}\label{P_thing}
P_{abcd} =& \,  \frac{\partial {\cal L}}{\partial R^{abcd}} \,, \nn\\
 =& \frac{1}{2 \kappa} g_{a[c}g_{b]d} + 6 \kappa \lambda \big[  \,  R _{ad} R _{bc} -  R_{ac} R _{bd} +  g_{bd} R _{a}{}^{e} \
R _{ce} -  g_{ad} R _{b}{}^{e} R_{ce}  -  g_{bc} R _{a}{}^{e} R_{de}  \nn\\
&+  g_{ac} R _{b}{}^{e} R_{de} 
-  g_{bd} R ^{ef} R_{aecf} +  g_{bc} R ^{ef} R _{aedf} + \
 g_{ad} R ^{ef} R _{becf} - 3 R_{a}{}^{e}{}_{d}{}^{f} R _{becf} \nn\\
&  - g_{ac} R ^{ef} R _{bedf} + 3 R_{a}{}^{e}{}_{c}{}^{f} R _{bedf} + \tfrac{1}{2} R_{ab}{}^{ef} R _{cdef} \big] \, .
\end{align}

As discussed in \cite{Bueno:2016xff}, remarkably, the linearized equations of this theory match those of Einstein gravity up to an overall constant.  Explicitly, if $h_{ab}$ is a perturbation away  from any maximally symmetric spacetime $\bar g_{ab}$ (i.e. $g_{ab} = \bar g_{ab} + h_{ab}$), then the linearized equations for $h_{ab}$ take the form,
\begin{align}
G^L_{ab} =&\, -\frac{1}{4 \kappa} \left(1 - \frac{96(D-3)(D-6) \kappa^2 \lambda \Lambda_{\rm eff}^2}{\left[(D-1)(D-2)\right]^2} \right) \bigg[\frac{4 \Lambda_{\rm eff} h_{ab}}{D -2} -  \frac{2 \Lambda_{\rm eff} \bar g_{ab} h^{c}{}_{c}}{D-2} + \nabla_{b}\nabla_{a}h^{c}{}_{c} \nn\\
&-  \nabla_{c}\nabla_{a}h_{b}{}^{c}  -  \nabla_{c}\nabla_{b}h_{a}{}^{c} + \nabla_{c}\nabla^{c}h_{ab} 
+ \bar g_{ab} \nabla_{d}\nabla_{c}h^{cd} -  \bar g_{ab} \nabla_{d}\nabla^{d}h^{c}{}_{c}\bigg]
\end{align}
where $\Lambda_{\rm eff}$ satisfies the equation\footnote{Note that our equation superficially differs from Eq. (25) of \cite{Bueno:2016xff}, where  the curvature of the maximally symmetric background is defined as $R_{abcd} = 2 \Lambda g_{a[c}g_{d]b}$. We instead have defined it to be $R_{abcd} = \frac{4 \Lambda_{\rm eff}}{(D-1)(D-2)} g_{a[c}g_{d]b}$.  The advantage of our method is that, when $\lambda=0$, our $\Lambda_{\rm eff}$ reduces to the standard cosmological constant, whereas the expression from \cite{Bueno:2016xff} coincides with what would more conventionally be termed $\pm 1/\ell^2$. },
\be\label{eff_cosmo} 
-\frac{32 (D-3)(D-6)}{\left[(D-1)(D-2) \right]^2} \kappa^2 \lambda \Lambda_{\rm eff}^3 + \Lambda_{\rm eff} - \Lambda = 0 \, .
\ee
It is now easy to ensure that the theory is free from ghosts by enforcing that the overall constant has the same sign as in the case of Einstein gravity, namely:
\be\label{ghost_free} 
1 - \frac{96(D-3)(D-6) \kappa^2 \lambda \Lambda_{\rm eff}^2}{\left[(D-1)(D-2)\right]^2} > 0 \, ,
\ee
which can be thought of as a constraint on $\lambda$.  We will come back to the issue of ghosts in the following subsection.

We now move on to consider black hole metrics in four-dimensions. Substituting a static, spherically symmetric metric of the form\footnote{A more general ansatz would include a lapse function. However the field equations force the lapse to be a constant, which can then be absorbed into the definition of $t$.  Thus there is no loss of generality in restricting to a metric of this form.}
\be 
ds^2 = -f(r) dt^2 + \frac{dr^2}{f(r)} + r^2 \left( d\theta^2 + \frac{\sin^2(\sqrt{k} \theta)}{k} d\phi^2 \right)
\ee
(where $k=-1,0,1$ corresponds to hyperbolic, planar, or spherical geometry for the $(t,r)~=~const.$ sector) leads to the following equation of motion,
\begin{align}\label{feq} 
G_r{}^r =& \, \frac{1}{2\kappa} \left[\frac{rf'+f-k + \Lambda r^2}{r^2} \right] + \lambda \kappa \left[ \frac{6 f f''' }{r^3} \left(r f' - 2f + 2k \right)  + \frac{6 f f''^2}{r^2} \right. 
\nn\\
&+ \left. \frac{24 f f''}{r^4} \left(f -k - rf' \right) +\frac{6 f'^2}{r^4} \left(4f - k\right) - \frac{24 f f'}{r^5} \left( f- k\right)\right] = 0\ 
\end{align}

The remaining components of the (generalized) Einstein tensor are either zero, equivalent to this expression, or equivalent to derivatives of this expression.  In the following three subsections we will consider perturbative solutions to this equation. 

\subsection{Asymptotic behaviour and ghosts}

We first examine the behaviour of the solution at large $r$.  To this end, we series expand the metric function in inverse powers of $r$,
\be 
f_{r \to \infty}(r) = k - \frac{\Lambda_{\rm eff}}{3} r^2 + \sum_n \frac{b_n}{r^n} \, ,
\ee
and substitute this series expansion into Eq.~\eqref{feq}.  Requiring the field equations to be satisfied results in the following series coefficients (with $b_1$ a free parameter),
\begin{align}\label{fhsol}
f_{r\to\infty}(r) =& \, k- \frac{\Lambda_{\rm eff}}{3}{r}^{2}+{\frac {b_{{1}}}{r}}-\,{\frac {168\Lambda_{\rm eff}\lambda\,{\kappa}^{
2}}{ \left( 16\,\Lambda_{\rm eff}^{2}{\kappa}^{2}\lambda+3 \right) }} \frac{b_1^2}{r^4} +\,{\frac {324\lambda\,{\kappa}^{2}k}{ \left( 16\,\Lambda_{\rm eff}
^{2}{\kappa}^{2}\lambda+3 \right) }} \frac{b_1^2}{r^6}
\nn\\
&+\,{\frac {12\lambda\,{
\kappa}^{2} \left( 20192\,\Lambda_{\rm eff}^{2}{\kappa}^{2}\lambda+69
 \right) }{ \left( 16\,\Lambda_{\rm eff}^{2}{\kappa}^{2}\lambda+3 \right) ^{2}}}  \frac{b_1^3}{r^7} 
-\,{\frac {1353024{\lambda}^{2}{\kappa}^{4}\Lambda_{\rm eff}k}{
 \left( 16\,\Lambda_{\rm eff}^{2}{\kappa}^{2}\lambda+3 \right) ^{2}}} \frac{b_1^3}{r^9}
 \nn\\
 &-\,{\frac {288{\lambda}^{2}{\kappa}^{4}\Lambda_{\rm eff} \left( 2881024\,\Lambda_{\rm eff}^{2}
{\kappa}^{2}\lambda+14457 \right) }{ \left( 16\,\Lambda_{\rm eff}^{2}{\kappa}^{2}
\lambda+3 \right) ^{3}}} \frac{b_1^4}{r^{10}}
\end{align}

Let us consider now the various possible asymptotics for these solutions.  In four dimensions, Eqs.~\eqref{eff_cosmo} and \eqref{ghost_free} become,
\begin{align}\label{4d_constraints}
\Lambda - \Lambda_{\rm eff} - \frac{16}{9} \lambda \kappa^2 \Lambda_{\rm eff}^3 = 0 \, , \nn\\
1 + \frac{16}{3} \lambda \Lambda_{\rm eff}^2 \kappa^2 > 0 \, .
\end{align}
Taking the discriminant of the first equation we find that,
\be 
\Delta = -\frac{64 \lambda}{9 \kappa^2} \left(1 + 12 \Lambda^2 \kappa^2 \lambda \right) \, .
\ee
The discriminant can be either positive, zero, or negative depending on the value of $\lambda$:
\begin{align}
 \Delta &> 0 \quad \text{ if } \,\, -\frac{1}{12 \Lambda^2 \kappa^2} < \lambda < 0 \, , \nn\\
 \Delta &= 0 \quad \text{ if } \,\, \lambda = 0 \,\, \text{ or } \,\, \lambda = -\frac{1}{12 \Lambda^2 \kappa^2} \, , \nn\\
 \Delta &< 0 \quad \text{ otherwise.}
\end{align}
\begin{figure}[htp]
\centering
\includegraphics[width = 0.5\textwidth]{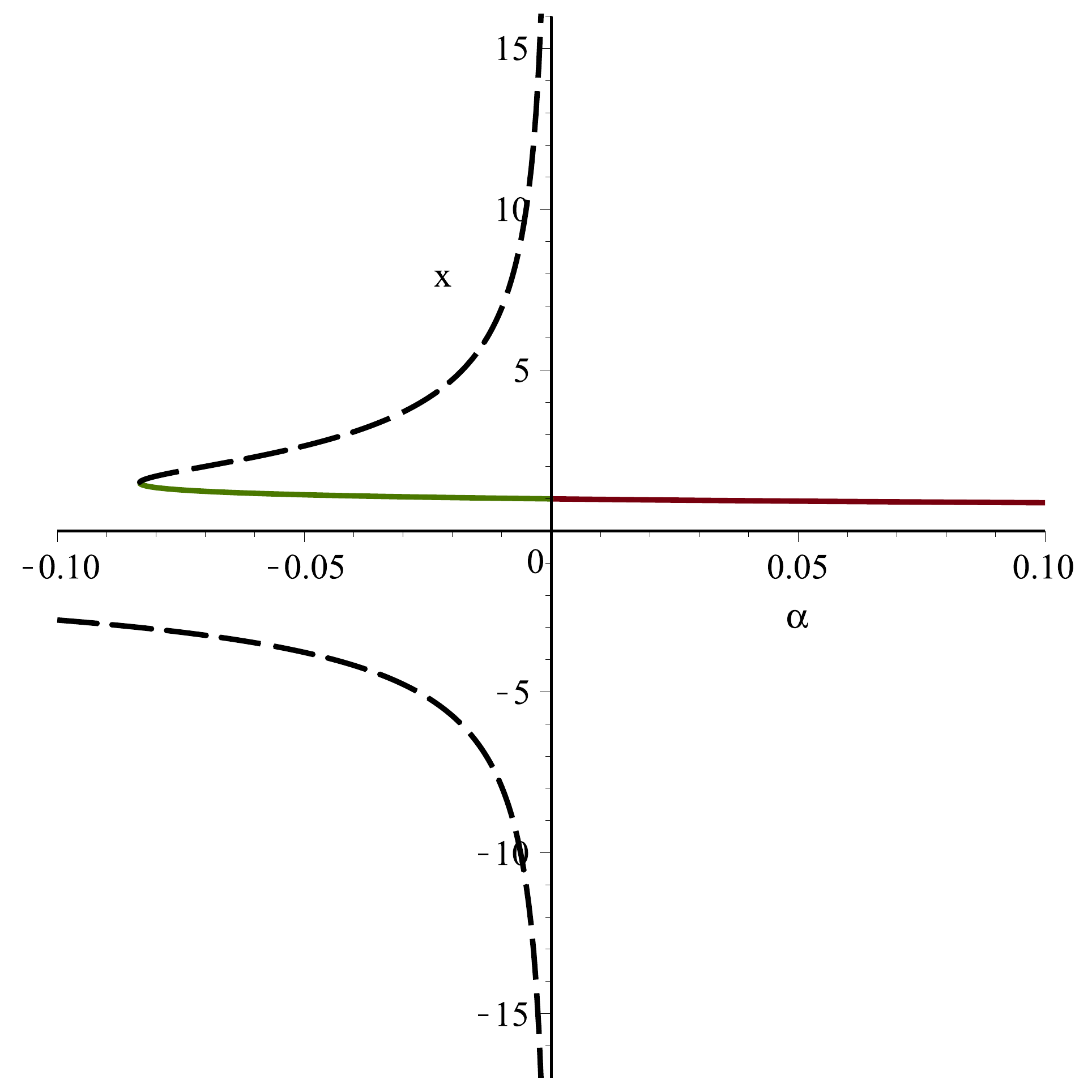}
\label{fig:ghost_condition}
\caption{{\bf Ghost condition}: This plot highlights the solutions for the effective cosmological constant vs. the cubic coupling where we have made use of the dimensionless parameters $x = \Lambda_{\rm eff}/\Lambda$ and $\alpha = \lambda \Lambda^2 \kappa^2$.  The dashed black lines indicate that the ghost-free condition is violated for these parameters.  We see that there is only ever a single ghost-free branch that  limits to the Einstein case when $\lambda \to 0$.  The fact that the ghost free branch has $x > 0$ indicates that the effective cosmological constant is of the same sign as the ordinary cosmological constant, $\Lambda$.  Here the green and red curves highlight that the branch that limits to Einstein gravity differs depending on whether $\lambda$ is positive or negative.}
\end{figure}
In the case $\Delta > 0$ the theory will have three real branches (i.e. three possible values for $\Lambda_{\rm eff}$), while it will have only one when $\Delta < 0$.  In the case $\Delta = 0$, the ghost condition is either trivial ($\lambda = 0$) or fails.  In general, we find that for any given $\Lambda$ and $\lambda$, there will only be a single branch that is ghost free, as shown in Figure~\ref{fig:ghost_condition}.  It is interesting to note that the ghost-free branch has a smooth limit to the Einstein case as $\lambda \to 0$; in other words, the Einstein branch is ghost free.  Furthermore, the ghost-free branch has the property that ${\rm sgn}(\Lambda) = {\rm sgn}(\Lambda_{\rm eff})$, meaning, for example, if $\Lambda < 0$ then the ghost-free branch will possess AdS asymptotics.

\subsection{Solution as a small $\lambda$ expansion}

We next consider treating the curvature cubed terms as a small correction to Einstein gravity.  Thus, we expand the metric function in terms of the dimensionless parameter $\lambda \Lambda^2$, which we treat as small,
\be 
f(r) = k - \frac{c_0}{r} - \frac{\Lambda r^2}{3} + \sum_{n=1} (\lambda\Lambda^2)^n h_n(r) \, .
\ee
Substituting this expression into the field equation yields a first order differential equation for $h_n(r)$ at each order in $\lambda$. These can be easily integrated to give the solution to arbitrary order in $\lambda$.  For example, solving for $h_1(r)$ and $h_2(r)$ gives,
\begin{align} 
h_1(r) &= \frac{\kappa^2}{\Lambda^2} \left[ {\frac {16\,{r}^{2}{\Lambda}^{3}}{27}} +{\frac {c_{{1}}}{r}} -\,{\frac {56\Lambda\,c_{
{0}}^{2}}{{r}^{4}}} +\,{\frac {108 kc_{{0}}^{2}}{{r}^{6}}} -\,{\frac {92c_{{0}}^{3}}{{r}^
{7}}} \right]
 \, , \nn\\
h_2(r) &= \frac{\kappa^4}{\Lambda^4} \left[ -{\frac {256\,{r}^{2}{\Lambda}^{5}}{81}}+{\frac {c_{{2}}}{r}}+ \left( {\frac {3584\,{\Lambda}^{3}c_{{0}}^{2}}{9}}+112\,
\Lambda\,c_{{0}}c_{{1}} \right) {\frac {
1}{{r}^{4}} }+{\frac {-576\,{\Lambda}^{2}kc_{{0}}
^{2}-216\,kc_{{0}}c_{{1}}}{{r}^{6}}} \right.
\nn\\ 
&+{ \left( -{
\frac {77824\,{\Lambda}^{2}c_{{0}}^{3}}{3}}+276\,c_{{0}}^{2}c_{{1}
} \right) }\frac {1}{{r}^{7}}+\,{\frac {150336\Lambda\,kc_{{0}}^{3}}{{r}^{9}}}-
\,{\frac {154208\Lambda\,c_{{0}}^{4}}{{r}^{10}}}-\,{\frac {217728{k}^{2}c
_{{0}}^{3}}{{r}^{11}}}
\nn\\
&+ \left. \,{\frac {443232kc_{{0}}^{4}}{{r}^{12}}}-
\,{\frac {224112c_{{0}}^{5}}{{r}^{13}}} \right] \, .
\end{align}
The higher order $h_n(r)$ terms can be obtained easily, but they are increasingly cumbersome and therefore not particularly illuminating.  

From this expansion (and the two representative terms shown above) we can see a few interesting properties.  First, since at each order the differential equation for $h_n(r)$ is first order, we have a single undetermined parameter $c_n$ at each order.  This parameter always appears as $c_n/r$ in the expansion, and therefore accounts for perturbative corrections to the mass parameter, $c_0$.  These contributions require suitable boundary conditions to fix: for example, one could demand that the horizon radius remains fixed and determine $c_n$ for $n>0$ in terms of the horizon radius. Furthermore, at each order there is a correction to the cosmological constant.  This is expected, since we know from the previous discussion that the higher curvature terms modify the asymptotics, effectively altering the cosmological constant.  Here what we are seeing is a perturbative expansion of the new, effective cosmological constant.  

\subsection{Near horizon solution \label{sec:4d_case_expansions}}

To study black hole solutions of this theory, it is useful to consider the metric function expanded near the horizon; this reads  
\be\label{nhsol}
f_{r \to r_+}(r) = \sum_{n=1} a_n (r-r_+)^n  
\ee
where the sum starts at $n=1$ since the metric function must vanish  linearly  for a non-extremal black hole.
Substituting this ansatz into the field equations produces a series of relationships that the coefficients $a_n$ must satisfy; for example the first two expressions are given by:
\begin{align}\label{eq:nhl}
0  = & \,  \Lambda r_+^4 + r_+^3 a_1 - k r_+^2 - 12 k \kappa^2 \lambda a_1^2 \, , \nn\\
0  = & \, (5\Lambda + 2 a_2) r_+^4 + \left(72 \kappa^2 \lambda a_1^2 a_3 + (5 + 48 \kappa^2 \lambda a_2^2)a_1\right)r_+^3 
\nn\\& \,+ (144 k \kappa^2 \lambda a_1 a_3 - 96 \kappa^2 \lambda a_1^2 a_2 - 3 k) r_+^2 - 48 \lambda a_1 \kappa^2 (3ka_2 - a_1^2) r_+ + 36 k \kappa^2 \lambda a_1^2 \, .
\end{align}
These two relationships suffice to highlight the general trend.  Notice that the first expression determines $a_1$, as we would expect.  There are in fact two solutions for $a_1$, one which has a smooth $\lambda \to 0$ limit, and another which does not. In the second expression both $a_2$ and $a_3$ appear, and this general trend continues: at order $(r-r_+)^n$, coefficients up to $a_{n+1}$ can occur.  The reason for this is the appearance of particular terms in the field equations involving third derivatives, e.g.
\be 
r^2 f(r) f'''(r) \, .
\ee

From here there are two courses of action one can follow.  First, we can treat $a_2$ as a free parameter, isolating the second relationship, i.e. contributions at ${\cal O}\left( (r-r_+)^2 \right)$, for $a_3$ in terms of $a_2$, and continuing this recursive procedure to higher orders.  This method produces two possible solutions due to the two initial choices of $a_1$.   However, regardless of which choice is made for $a_1$ neither of these solutions has a smooth $\lambda \to 0$ limit, that is, neither is the Einstein branch.  This is true irrespective of the choice of $a_2$; that is, while $a_2$ may be selected to cancel a divergence\footnote{ Here we mean a divergence in the small $\lambda$ limit:  as $\lambda \to 0$, $a_n \to \infty$.} in some particular $a_n$, this choice of $a_2$ will not cure the divergences in all of the coefficients.   Thus this procedure allows us to study the two non-Einsteinian cubic branches of the theory.  However, as discussed earlier, these branches both suffer from the ghost instability.

To obtain information about the Einstein branch, we have found it necessary to follow a second approach, which we shall now describe.  Specifically, we consider the small $\lambda$ limit and work perturbatively, expanding each of the $a_n$ series coefficients above in powers of $\lambda$:
\be 
a_n = \sum_{j=0}^{j_{\rm max}} b_{n, j} \lambda^j \, . 
\ee
Using this expression in the field equations, we find that to obtain a series that is accurate to ${\cal O}(\lambda^n)$ at ${\cal O}\left( (r-r_+)^m \right)$ it is necessary to work to ${\cal O}(\lambda^{n-1})$ at ${\cal O}\left( (r-r_+)^{m+1} \right)$ and so on until ${\cal O}(\lambda^{0})$ at ${\cal O}\left( (r-r_+)^{m+n} \right)$.  As expected, the $b_{n, 0}$ terms reproduce the series expansion of the ordinary Einstein equation about an event horizon,
\begin{align}
b_{1,0} &= \frac{k-\Lambda r_+^2}{r_+} \, , \quad b_{2, 0} = -\frac{k}{r_+^2} \, , \quad b_{3,0} = \frac{3k - \Lambda r_+^2}{3 r_+^3} \, , \quad b_{n,0} = (-1)^{n+1} \frac{b_{3,0}}{r_+^{n-3}} \, .
\end{align} 
The first and higher order terms in $\lambda$ do not follow a nice pattern, and so we present here only some sample coefficients.  Explicitly, taking $j_{\rm max} = 2$ 
and requiring ${\cal O}(\lambda^2)$ accuracy at ${\cal O} \left( (r-r_+)^3 \right)$ we find the following coefficients:
\begin{align}
b_{1, 1} &= \frac{12 k \kappa^2 (\Lambda r_+^2 - k)^2}{r_+^5} \, , \quad b_{1,2} = -\frac{288 k^2 \kappa^4(\Lambda r_+^2 -k )^3}{r_+^9} \, , 
\nn\\
b_{2,1} &= \frac{36 \kappa^2 (\Lambda r_+^2 - k)(\Lambda r_+^2 - 3 k)^2}{r_+^6} \, , 
\nn\\
b_{2,2} &= \,{\frac { 144 \left( \Lambda\,r_+^{2}-3\,k \right) {
\kappa}^{4} \left( \Lambda\,r_+^{2}-k \right) }{r_+^{10}} \left( 53{
\Lambda}^{3}r_+^{6}- 392 \Lambda^2 r_+^4 k + 879 \Lambda r_+^2 k^2 - 564 k^3 \right) 
}\, ,
\nn\\
b_{3,1} &= -\,{\frac {4\,{\kappa}^{2} \left( 41\,{\Lambda}^{3}r_+^{6}-294
\,{\Lambda}^{2}r_+^{4}k+653\,\Lambda\,r_+^{2}{k}^{2}-424\,{k}^{3}
 \right) }{r_+^{7}}} \, ,
 \nn\\
b_{3,2} &= -\frac{48}{r_+^{11}} \big[\,{{\kappa}^{4} \left( 1801\,{\Lambda}^{5}r_+^{
10}-20526\,{\Lambda}^{4}r_+^{8}k+89460\,{\Lambda}^{3}r_+^{6}{k}^{2}-
184398\,{\Lambda}^{2}r_+^{4}{k}^{3} \right. } 
\nn\\
&+177787\, \left. \Lambda\,r_+^{2}{k}^{4}-
64220\,{k}^{5} \right) \big] \, .
\end{align}

Working to higher order, we can show that,
\be 
a_1 = \sum_{j=0}^{\infty} b_{1, j} \lambda^j = \frac{r_+}{24  \lambda k \kappa^2} \left[r_+^2 - \sqrt{48k \kappa^2 (\Lambda r_+^2 - k) \lambda + r_+^4} \right] \, ,
\ee
which is the Einsteinian root for $a_1$ from Eq.~\eqref{eq:nhl}. We have not been able to find corresponding closed-form expressions for $a_2$ and $a_3$ by summing the series. 
As we shall see later, this expression for $a_1$ is all that is needed to characterize the black hole thermodynamics of the Einstein branch.

\subsection{Thermodynamics \label{sec:4d_limits}}

We now move on to thermodynamic considerations for these black hole solutions given by
\eqref{fhsol} and \eqref{nhsol}.  Their temperature can be computed by requiring regularity of the Euclidean sector, giving
\be 
T = \frac{f'(r_+)}{4 \pi} = \frac{a_1}{4 \pi}
\ee
where $a_1$ solves the equation
\be\label{a1_eq} 
\Lambda r_+^4 + r_+^3 a_1 - k r_+^2 - 12 k \kappa^2 \lambda a_1^2 = 0 \, .
\ee
The entropy can be calculated using Wald's prescription \cite{Wald:1993nt, Iyer:1994ys},
\be 
S = -2 \pi \oint d^2 x \sqrt{\sigma} P^{abcd} \hat{\epsilon}_{ab} \hat{\epsilon}_{cd}
\ee
with $\sigma$ the determinant of the induced metric on the horizon and $\hat{\epsilon}_{ab}$ the binormal to the horizon, $\hat \epsilon_{ab} \hat \epsilon^{ab} = -2$.  Using $P^{abcd}$ as defined above, we find that the Wald entropy is given by
\be\label{eq:entropy} 
S = \frac{2 \pi A}{\kappa} \left[1 + \frac{12 \lambda \kappa^2 a_1}{r_+^3} \left(r_+ a_1 + 4k\right) \right] \, ,
\ee
which, remarkably, depends only on $a_1$.  Finally, as is standard in black hole chemistry, we take the pressure to be given by 
\be 
P = -\frac{\Lambda}{\kappa}
\ee
for AdS black holes.  We shall return to the full thermodynamics after some relevant discussion of the series coefficient $a_1$.  

The equation determining $a_1$ (Eq.~\eqref{a1_eq}) is quadratic in $a_1$, with the two solutions (assuming $k = \pm 1$) being
\begin{align}
a_1^\pm = \frac{r_+}{24  \lambda k \kappa^2} \left[r_+^2 \pm \sqrt{48k \kappa^2 (\Lambda r_+^2 - k) \lambda + r_+^4} \right] \, .
\end{align}  
Note that $a_1^-$ has a smooth $\lambda \to 0$ limit, while $a_1 ^+$ does not.  However, there are some consistency conditions  and additional restrictions we must consider if we wish to take the small $\lambda$ limit.  The considerations will be important for the higher dimensional cases we consider in the next section, and are most easily understood here in the four dimensional case.

For convenience, we employ the dimensionless parameters
\be 
\alpha = \kappa^2 \Lambda^2 \lambda \, , \quad \tilde r_+ = \sqrt{-\frac{3}{\Lambda}} r_+ \, ,
\ee
in terms of which, $a_1^{\pm}$ takes the form,
\be 
a_1^\pm = \tilde r_+^2 \sqrt{\frac{-\Lambda}{3}} \left\{ \frac{1}{8 \alpha k \tilde r_+} \left[ 3 \tilde{r}_+^2 \pm \sqrt{9 \tilde r_+^4 - \alpha \left(48 k^2 + 144 k \tilde r_+^2 \right)} \right] \right\}
\ee
First, note that the term under the square root could become negative.  To prevent this, we must demand that
\be\label{horcond}
9 \tilde r_+^4 - \alpha \left(48 k^2 + 144 k \tilde r_+^2 \right) > 0 \, .
\ee  
This equation is trivially positive in the case $k=0$, and so we restrict our attention to the cases $k=\pm 1$. For $k=+1$, we find that if $\alpha > 0$, then the minimum horizon radius is given by
\be 
\tilde r_{+, min}^{k=+1} = \sqrt{8\alpha + \frac{4}{3} \sqrt{3\alpha(12\alpha+1)}}
\ee 
and for $\alpha < 0$ any horizon radius is permitted.  If $k=-1$, the situation is slightly more complicated.  If $\alpha > 0$, then for fixed $\alpha$, there is a minimum horizon size given by
\be 
\tilde r_{+, min}^{k=-1} = \frac{2}{3} \sqrt{3\sqrt{3\alpha(12 \alpha+1)} - 18 \alpha} \, \quad \text{for } \alpha > 0 \, .
\ee  
When $-1/12 < \alpha < 0$ there are no restrictions on the horizon radius; \eqref{horcond} is always satisfied.   However for $\alpha < -\frac{1}{12}$ there are no black holes for $r_+$ satisfying
\be 
\frac{2}{3} \sqrt{3\sqrt{3\alpha(12 \alpha+1)} - 18 \alpha} < r_+ < \frac{2}{3} \sqrt{-3\sqrt{3\alpha(12 \alpha+1)} - 18 \alpha} \quad \textrm{for } \alpha < -\frac{1}{12}\,.
\ee

We get further constraints upon considering the small coupling limit. If we wish to expand $a_1^{-}$ as $\lambda \to 0$ (i.e. $\alpha \to 0$), we are performing a Taylor expansion of the square root and therefore must, in addition to the above constraints, have,
\be 
\left| \frac{\alpha \left(48 k^2 + 144 k \tilde r_+^2 \right)}{9 \tilde r_+^4}\right| < 1 \, .
\ee
These conditions taken together imply a minimum black hole size in the small coupling limit, and we can solve for this exactly.  We can express  the value for $\alpha$ which yields $r_{+, min}$ for the minimum horizon size concisely as
\be\label{4d_small_lambda_constraint} 
|\alpha | =  \frac{3}{16} \frac{\tilde r_{+,min}^4}{k(k+3 \tilde r_{+,min}^2)}
\ee  
where it is understood that only positive values of $\tilde r_{+,min}$ are permissible;  the absolute value bars indicate that $\tilde r_{+,min}$ is the same, regardless of the overall sign of $\alpha$.  Note that for $k=-1$, the above has a pole for $\tilde r_{+,min} = \sqrt{-k/3}$.  This corresponds to the value of $\tilde r_+$ that causes the coefficient of $\alpha$ under the square root in the expression for $a_1$ to vanish.  For this special value of $r_+$, the $\alpha$ (and therefore $\lambda$) dependence drops out of the square root, and no small coupling limit exists for $a_1$.

\subsubsection{Thermodynamics of Einstein branch}

We shall now study in more detail the thermodynamics of black hole solutions of the Einstein branch.  As discussed earlier, this is the only branch that is free from the ghost instability, i.e. the only branch with a well defined ground state.  For the Einstein branch, 
\be 
a_1 = \frac{r_+}{24  \lambda k \kappa^2} \left[r_+^2 - \sqrt{48k \kappa^2 (\Lambda r_+^2 - k) \lambda + r_+^4} \right] \, .
\ee
The first law and Smarr formula,
\begin{align}
dM &= TdS + V dP + \Psi_\lambda d\lambda \, , \nn\\
M &= 2(TS-VP) + 4 \Psi_\lambda \lambda \, ,
\end{align}
are satisfied by the thermodynamic potentials defined in the previous section along with the following identifications for the mass, volume and conjugate to $\lambda$:
\begin{align}
M =& \, \frac{ \pi r_+^3}{216 k^3 \kappa^5 \lambda^2} \left[r_+^6 - r_+^4\left(Y - 36 k \kappa^2 \lambda \Lambda \right) - 12 k \kappa^2 \lambda \Lambda r_+^2 Y + 24 k^2 \kappa^2 \lambda \left(24 k \kappa^2 \lambda \Lambda - Y \right) \right] \, , \nn\\
Y =& \, \sqrt{48k \kappa^2 (\Lambda r_+^2 - k) \lambda + r_+^4} \, , \nn\\
V =& \,  \frac{4}{3} \pi r_+^3 \, , \nn\\
\Psi_\lambda =& \,  \frac{1}{4\lambda} \left[M - 2(TS -VP) \right]
\end{align}

In the limit of small $\lambda$, the mass takes the form,
\be 
M_{\lambda \to 0} = -\frac{4 \pi r_+}{3 \kappa} (r_+^2 \Lambda - 3k) + \frac{16 \pi \kappa \left( k - r_+^2 \Lambda \right)^2(4k-r_+^2\Lambda)}{r_+^3}  \lambda + {\cal O}(\lambda^2)
\ee
where we see that the leading order term is the standard Schwarzschild-(A)dS mass.   From the second term of the expansion, one might be tempted to think that the mass of the black hole tends to infinity as $r_+ \to 0$.  However, this is not the case: as  discussed earlier, for a given fixed $\lambda$, there is a minimum value for $r_+$.  Hence $r_+$ cannot be taken directly to zero in this expansion, but only to the minimum value specified by Eq.~\eqref{4d_small_lambda_constraint}.

We can study   $P-v$ criticality by constructing the equation of state.  Rearranging the definition of temperature and setting $\kappa = 8 \pi$, we obtain for the pressure,
\be 
P = \frac{T}{2 r_+} - \frac{k}{8 \pi r_+^2} - \frac{24 \pi k (8 \pi)^2 \lambda T^2}{r_+^4} \, .
\ee
It is easy to see that in the Einstein limit ($\lambda \to 0$), the last term drops out and the resulting equation of state possesses no inflection points. This is a qualitatively different equation of state than has appeared in Einstein, Lovelock, and quasi-topological theories of gravity for spherically symmetric black holes as it is quadratic (and not just linear) in $T$.
As a result of the cubic curvature corrections, this equation of state admits a single critical point provided $k=1$ and $\lambda < 0$ with critical values
\be 
P_c = \frac{\sqrt{2}}{256 \pi (8 \pi) \sqrt{-\lambda}} \, , \quad 
T_c = \frac{2^{1/4}}{12 \pi \sqrt{8 \pi} (-\lambda)^{1/4}} \, , \quad
r_c = 2^{7/4} \sqrt{8 \pi} (-\lambda)^{1/4} \, .
\ee
The ratio of these critical values exactly matches that of the van der Waals fluid
\be 
\frac{P_c v_c}{T_c} = \frac{P_c (2r_c)}{T_c} = \frac{3}{8}
\ee
which is independent of the parameters of this black hole solution, and is therefore a universal quantity for black holes of this theory.   The only other time this same ratio has been seen is for the Reissner Nordstrom black hole \cite{Kubiznak:2012wp}. The reason is that both solutions   have the same falloff behavior in $r_+$, namely $1/r^4_+$. The additional factor of $T^2$ doesn't make any difference because one computes derivatives with respect to $r_+$. 
  
To better understand the critical behaviour of these black holes we introduce the following dimensionless parameters:
\be
p = \sqrt{-\lambda} P \, , \quad v = 2 (-\lambda)^{-1/4}  r_+ \, , \quad t= (-\lambda)^{1/4} T \, ,
\ee
which allow us to study the thermodynamic behaviour for $\lambda < 0$.  In terms of these quantities, the equation of state reads,
\be 
p_{\lambda < 0} = \frac{t}{v} - \frac{k}{2 \pi v^2} + \frac{384 \pi k (8 \pi)^2 t^2}{v^4}
\ee
which, it must be kept in mind, applies only for $\lambda < 0$.  In terms of these dimensionless quantities, the critical point occurs at the values,
\be 
p_c = \frac{\sqrt{2}}{2048 \pi^2 } \, , \quad v_c = 4\, 2^{3/4} \sqrt{8 \pi} \, , \quad t_c = \frac{2^{1/4}}{12 \pi \sqrt{8 \pi}} \, . 
\ee
Expanding the equation of state about the critical point in terms of the following parameters:
\be\label{eqn:near_critical} 
\rho = \frac{p}{p_c} - 1 \, , \quad \omega = \frac{v}{v_c} - 1 \, , \quad \tau = \frac{t}{t_c} - 1 \, ,
\ee
we find that it takes the form,
\be 
\rho = \frac{10}{3} \tau  - \frac{4}{3} \tau \omega - \frac{4}{3} \omega^3 + {\cal O}(\tau \omega^2, \omega^4) \, .
\ee
From this expansion it is straightforward to show (see, e.g. \cite{Gunasekaran:2012dq}) that the critical exponents are given by the mean field theory values,
\be 
\alpha = 0 \, , \quad \beta = \frac{1}{2} \, , \quad \gamma = 1 \, , \quad \delta = 3 \, .
\ee
These four parameters characterize the specific heat at constant volume, order parameter, isothermal compressibility, behaviour of pressure along the critical isotherm, respectively, as the critical point is approached.  Interestingly, we have obtained the results suggested by~\cite{Majhi:2016txt, Mandal:2016anc} despite having a different form for the equation of state, with terms quadratic in the temperature.  

To determine if there are any phase transitions, we study the Gibbs free energy of these black holes,
\be 
G = M - TS \, ,
\ee
which can be converted to dimensionless form via the dimensionless parameters defined above and the rescaling,
\be 
g = (-\lambda)^{1/4} G \, .
\ee
\begin{figure}[htp]
\centering
\includegraphics[width=0.3\textwidth]{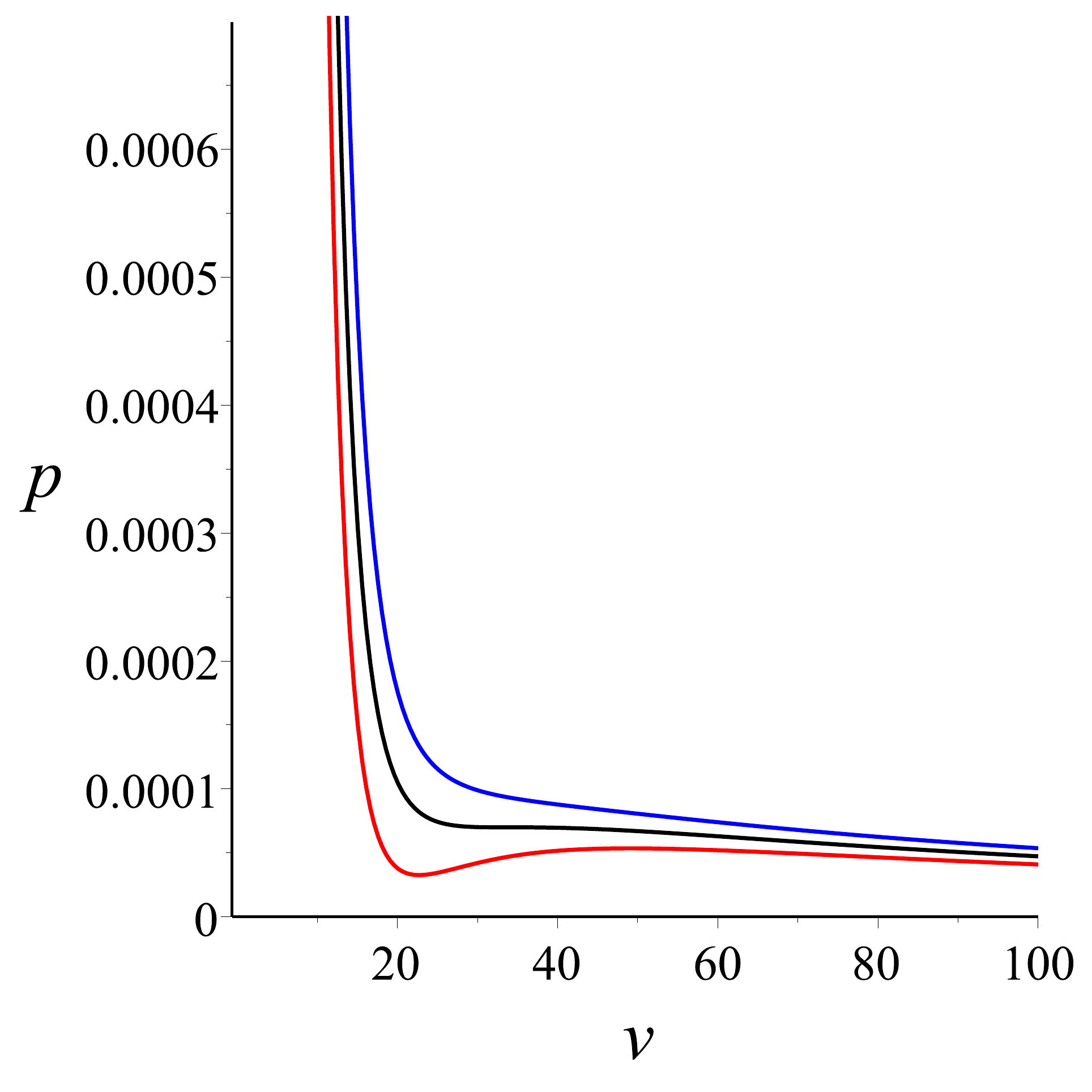}
\includegraphics[width=0.3\textwidth]{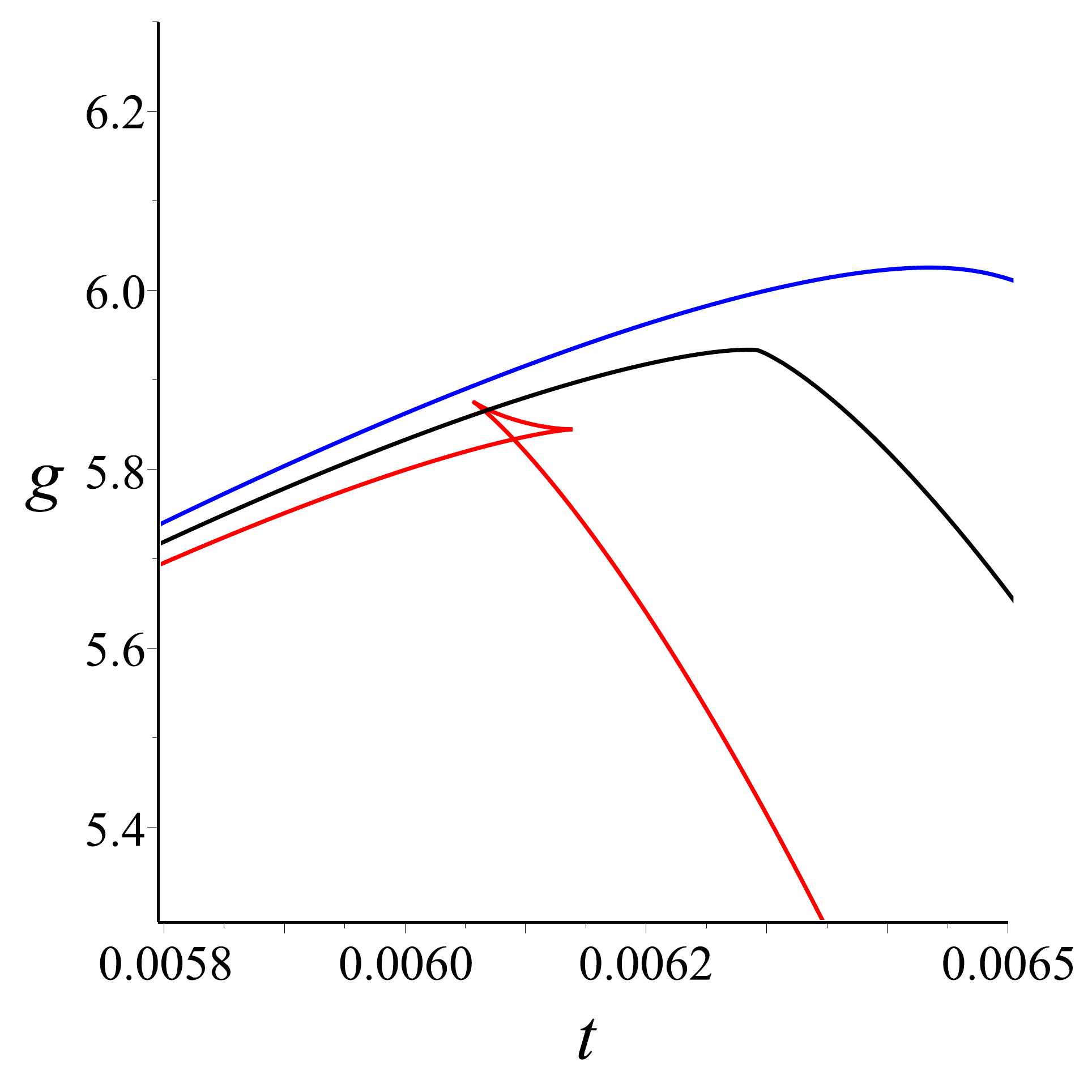}
\includegraphics[width=0.3\textwidth]{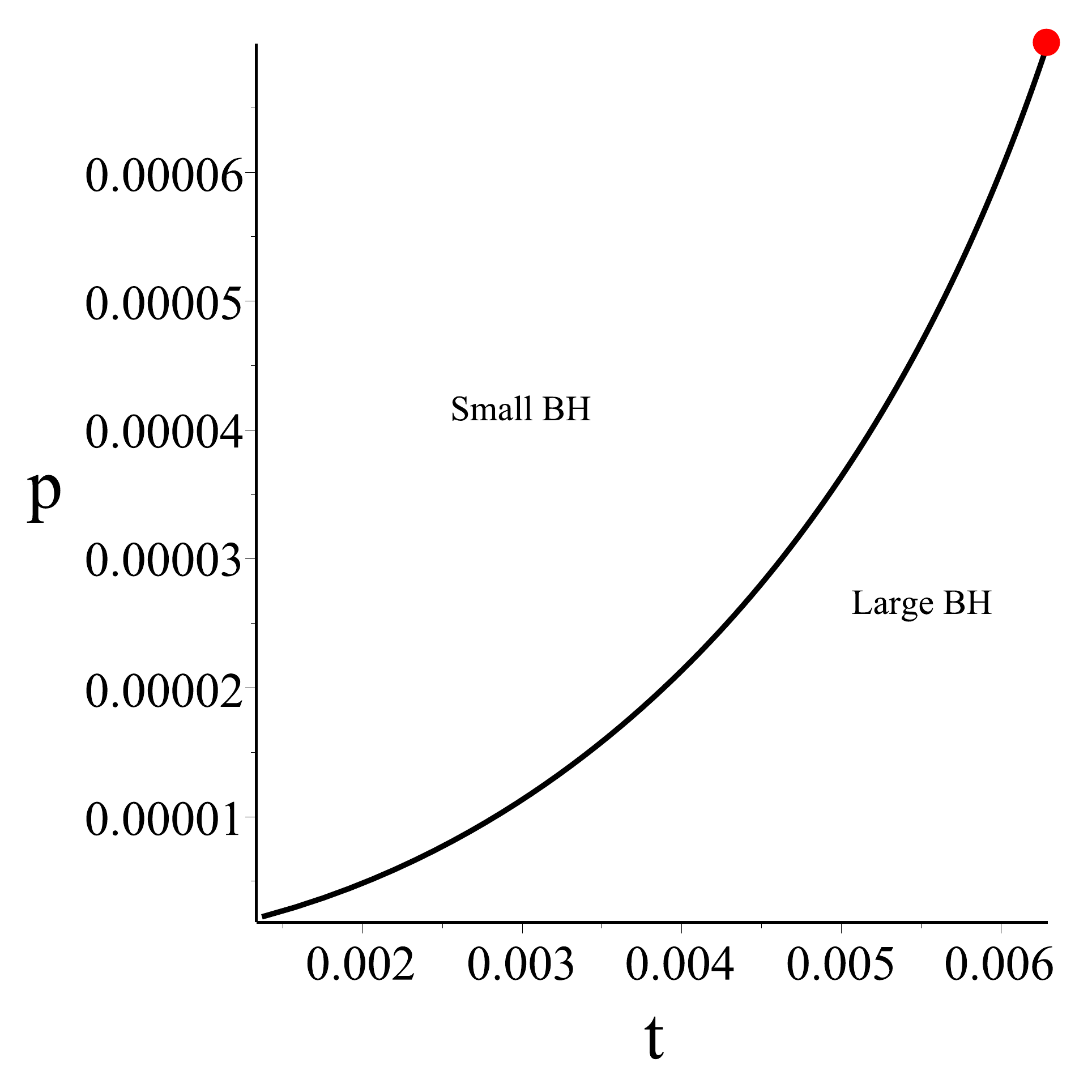}
\caption{{\bf Criticality for $\lambda < 0$}: $k=1$, $\kappa = 8\pi$:  {\it Left}: A $p-v$ plot showing van der Waals behaviour for $t=0.9t_c,\, t_c,\, 1.1t_c$ (bottom to top).  {\it Center}: A plot of the Gibbs free energy vs. temperature showing swallowtail behaviour for $p< p_c$.  The curves correspond to $p=0.9p_c,\, p_c,\, 1.1p_c$ from bottom to top. {\it Right}: The critical behaviour in the $p-t$ plane.  Here we see typical van der Waals behaviour: the black coexistence line separates small/large black hole phases until it terminates at the critical point, illustrated in this plot by a red dot.   Note that no constraints have been enforced on the entropy in the construction of these plots.  }
\label{fig:critical_no_entropy}
\end{figure}

\begin{figure}[htp]
\centering
\includegraphics[width=0.4\textwidth]{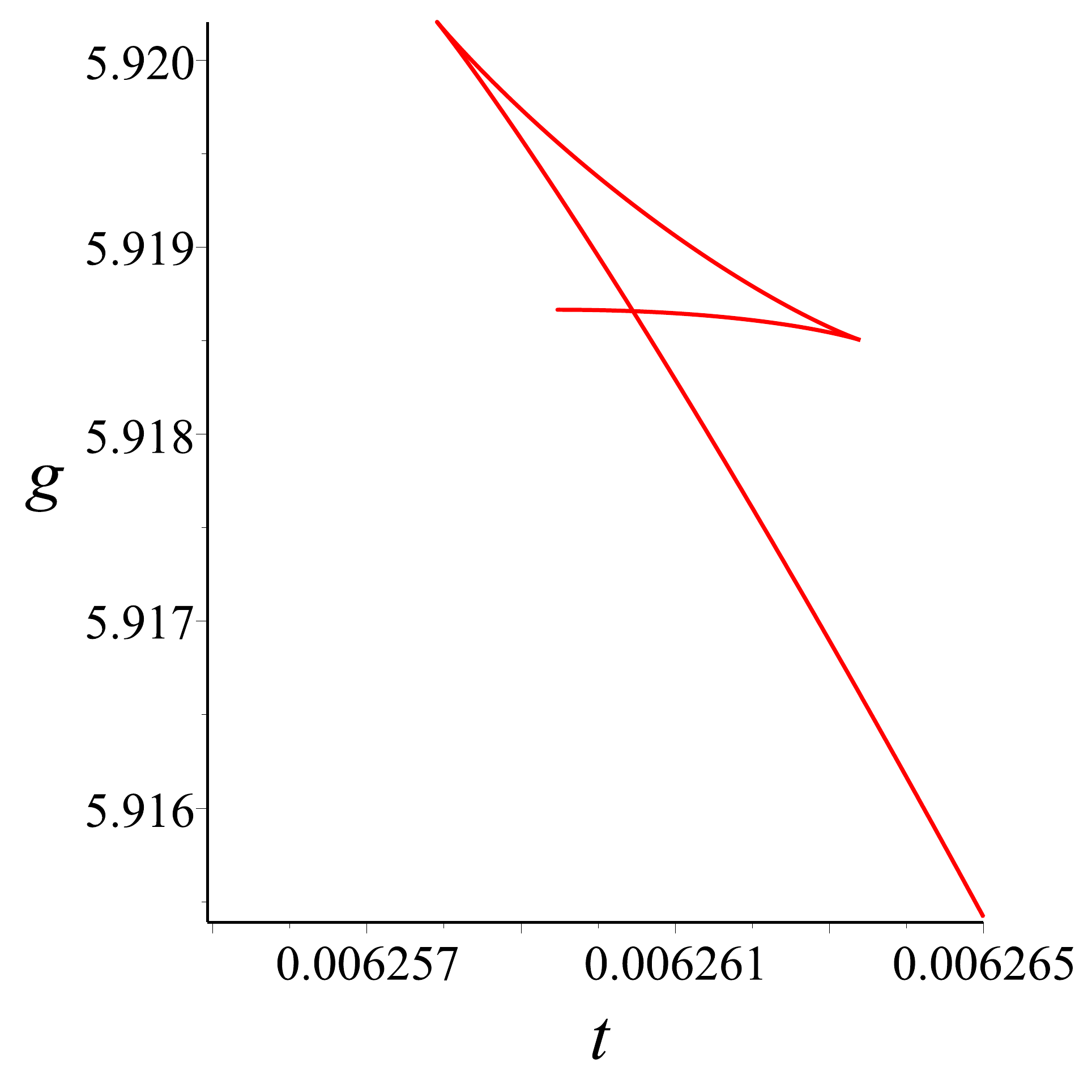}
\includegraphics[width=0.4\textwidth]{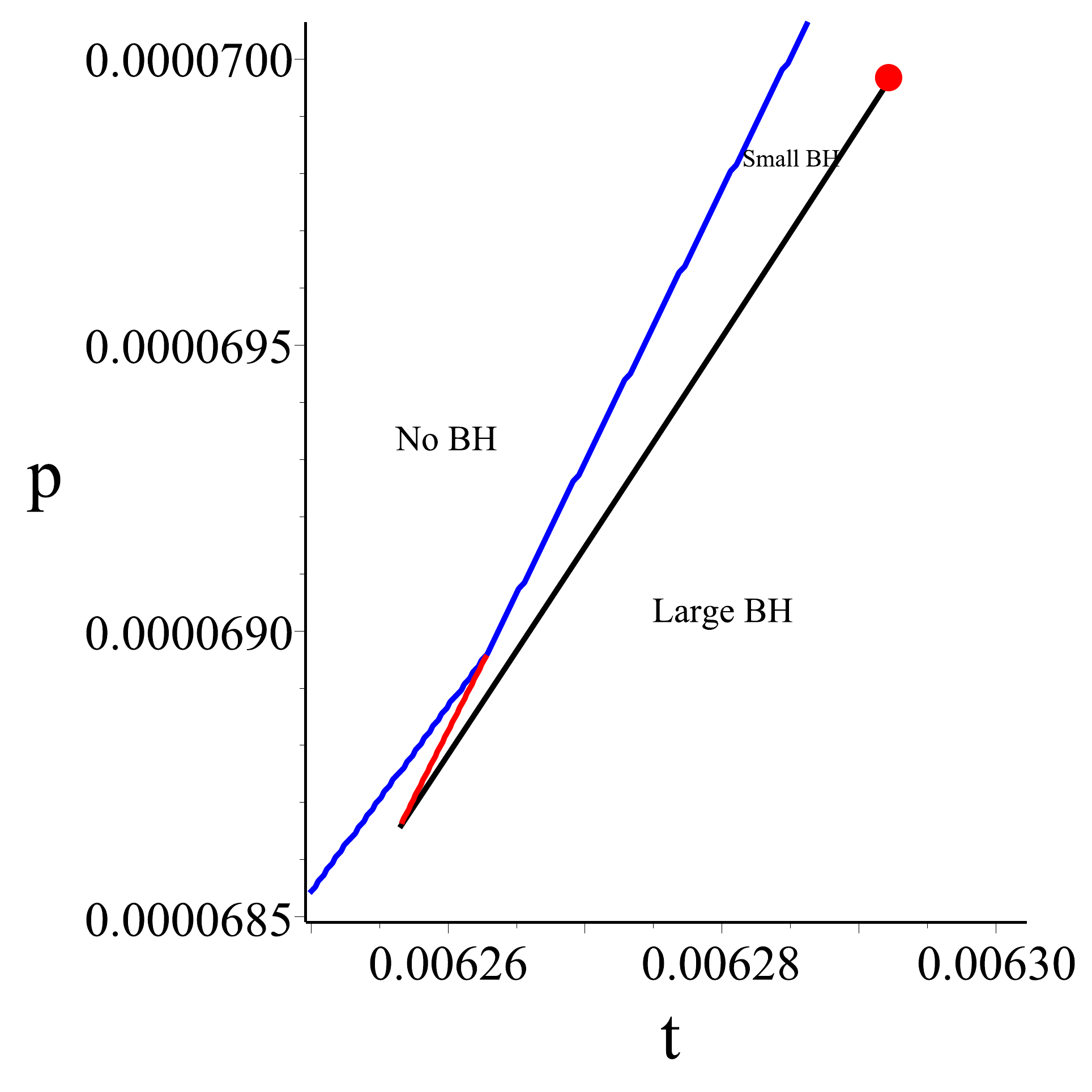}
\caption{{\bf Criticality for $\lambda < 0$}: $k=1$, $\kappa = 8\pi$: These plots of the Gibbs free energy vs. temperature (left) and $p$ vs. $t$ (right) illustrate how the plots from Figure~\ref{fig:critical_no_entropy} are altered if one demands positivity of entropy as a physical constraint.  The van der Waals type behaviour is replaced by a large/small/large black hole reentrant phase transition over a very small window of pressures.  A ``no black hole'' region is introduced where, for parameters in this region, the theory does not have positive entropy black hole solutions. The plot of the Gibbs free energy is for $p = 0.0000688$.  }
\label{fig:critical_entropy}
\end{figure}

The state of the physical system is taken to be that which minimizes the Gibbs free energy at constant temperature and pressure. Exploring first the behaviour near the critical point, we find that these black holes display van der Waals type behaviour, with the liquid/gas phase transition replaced by a small/large black hole phase transition.  Representative plots are shown in Figure~\ref{fig:critical_no_entropy}, where we see the standard van der Waals-type oscillation in the $p-v$ plane, swallowtail behaviour in the Gibbs free energy, and in the $p-t$ plane a coexistence line of a first order phase transition terminating at the critical point.  We note from the plots of the Gibbs free energy that the two physical branches of the Gibbs free energy both possess positive specific heat, therefore ensuring that any black hole which is a local minimum of the Gibbs free energy is also (locally) thermodynamically stable.   

Since we are considering higher curvature gravity, care must be taken when dealing with the black hole entropy --- in the case of $\lambda < 0 $ and $k=1$, we can see from Eq.~\eqref{eq:entropy} that it is possible for the entropy to be negative.  Assuming that the Wald entropy correctly identifies the true entropy of the black hole, and furthermore that this entropy is related to the underlying microscopic degrees of freedom,  negativity of the entropy 
should be regarded as unphysical.  However the negative entropy solutions of the theory do not seem to be affected by any other pathology.  Thus one could argue that these black holes are in fact physical and perhaps the Wald entropy should be supplemented by the addition of some positive, arbitrary constant.  

Due to this inherent ambiguity, we have considered both points of view. Noting that 
the plots in Figure~\ref{fig:critical_no_entropy} do not impose any constraints on the entropy, we plot 
in Figure~\ref{fig:critical_entropy} the relevant thermodynamic quantities imposing the positive entropy constraint.    Here we see that the van der Waals type behaviour has been modified via the addition of a ``no black hole" region, which corresponds to parameter values for which there are no positive entropy black holes of the theory.  We also see the appearance of a zeroth order phase transition connecting the first order coexistence line and the no black hole region.  This gives rise to a large/small/large reentrant phase transition for a very small window of pressures.  The first order coexistence line terminates at the critical point.   
\begin{figure}
\centering
\includegraphics[width=0.4\textwidth]{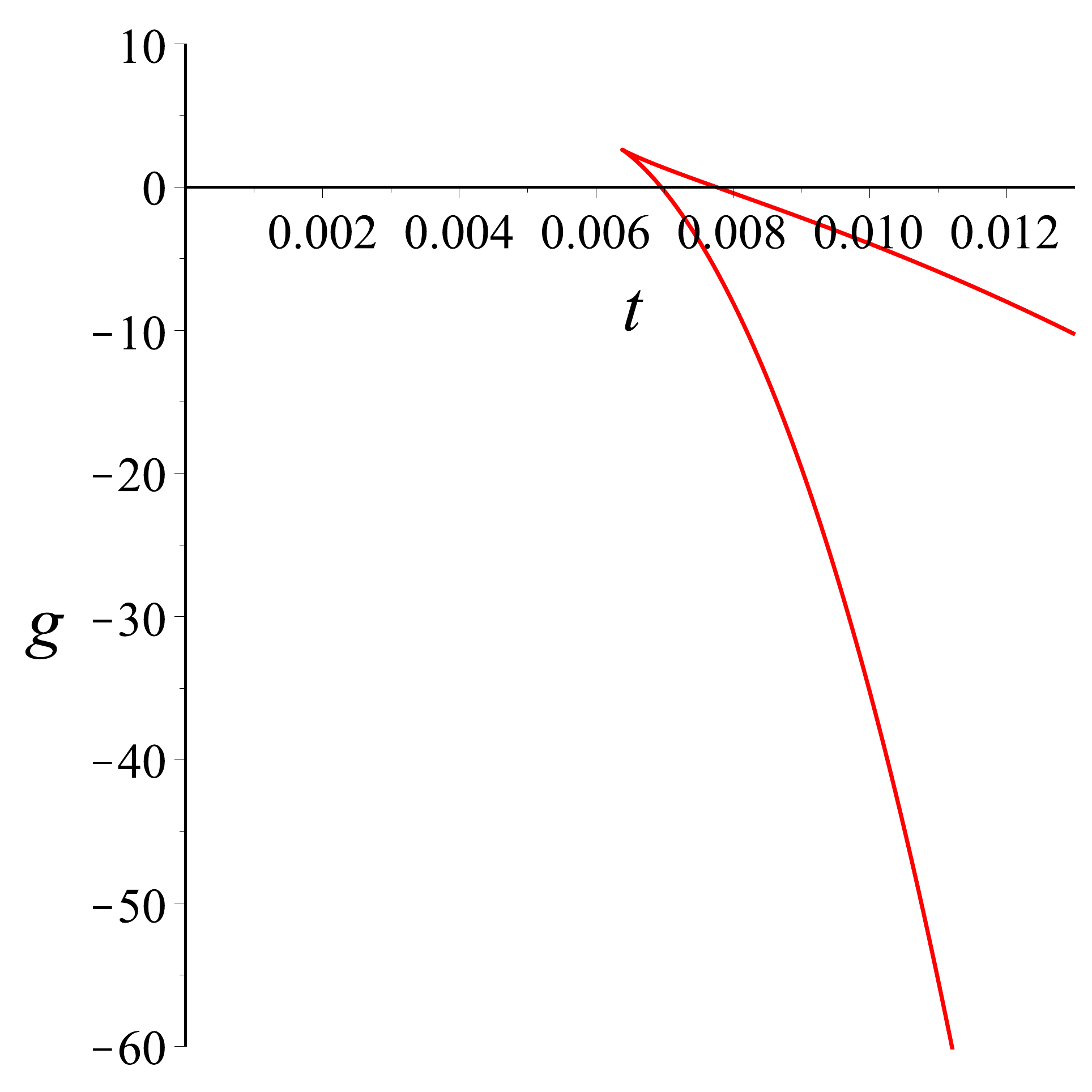}
\caption{{\bf Plot of Gibbs free energy}: $\lambda >0$, $k=1$, $\kappa = 8 \pi$, and $p=0.00006$.  This plot shows a representative example of the form of the Gibbs free energy when $(k<0, \lambda>0$), $(k >0, \lambda>0)$, $(k < 0, \lambda > 0)$ where it generally takes the form of a cusp, and no critical behaviour is observed. }
\label{fig:gibbs_representative}
\end{figure}

This concludes the discussion of the thermodynamics in the case of $\lambda < 0$ and $k=1$.  Although there are clearly three more combinations of these parameters that could can be studied, for each  the Gibbs free energy takes the form of a cusp, as shown in Figure~\ref{fig:gibbs_representative}. 
  Hence we do not find any instances of critical behaviour in these cases.

\section{Solutions in higher dimensions \label{sec:higherD}}

We now move on to considering solutions in higher dimensions.  It is advantageous to work with dimensionless coordinates; we therefore take as the metric ansatz the following:
\be\label{higherDmetric} 
ds^2 = \ell^2 \left[-\tilde r^2 N^2(\tilde r )f(\tilde r) d \tilde{t}^2 + \frac{d \tilde r^2}{\tilde r^2 f(\tilde r)} +  {\tilde r}^2 d\Sigma^2_{(k) D-2} \right]
\ee
where $d\Sigma^2_{(k)D-2}$ is the line element on a $(D-2)$-dimensional surface of constant scalar curvature $k (D-2)(D-3)$ and $k = +1,0,-1$ describes spherical, flat and hyperbolic geometries, respectively as before; in the latter cases the space can be made compact via appropriate identifications~\cite{Mann:1997iz}.  In the above, we have identified the cosmological constant as
\be
\Lambda = -\frac{(D-1)(D-2)}{2 \ell^2}  
\ee
and $\tilde t= t/\ell$ and $\tilde r = r/\ell$ to be dimensionless quantities; we have also pulled a factor of $\tilde r^2$ out of $f(\tilde r)$ so that asymptotically $f(\tilde r) \to 1$.

We proceed to construct the solution as a small $\lambda$ series, expanding the metric functions as
\begin{align}
f(\tilde r) &= 1 + \frac{k}{\tilde r^2} - \frac{c_0}{\tilde r^{D-1}} + \sum_{i=1}^{i_{\max}} \alpha^i h^{(D)}_i(\tilde r) \, , 
\nn\\
N(\tilde r) &= 1 + \sum_{i}^{i_{\max}} \alpha^i N_i^{(D)}(r)  
\end{align}
where $\alpha := \lambda \kappa^2 \Lambda^2$ is a dimensionless parameter.  Evaluating the field equations,
\be 
G_{a b} = P_{a cde}R_{b}{}^{cde} - \frac{1}{2} g_{ab} {\cal L} - 2 \nabla^c \nabla^d P_{acdb} = 0
\ee
($P_{a cde}$ is the same as given in Eq.~\eqref{P_thing}) on this ansatz produces 
\begin{align}\label{fo_correction}
h^{(D)}_1(\tilde r) =&  \frac{48 (D-3)(D^3 - 4 D^2 + 7D - 14)}{(D-2)^2(D-1)^2} \frac{c_0^2}{\tilde r^{2(D-1)}} + \frac{48 k (D-3)^2}{(D-2)^2} \frac{c_0^2}{\tilde r^{2D}}
\nn\\
&- \frac{4(D-3)(7D^3 - 30 D^2 + 47 D - 64)}{(D-1)^2(D-2)^2} \frac{c_0^3}{\tilde r^{3(D-1)}}  + \frac{c_1}{\tilde r^{D-1}} + \frac{32 (D-3)(D-6)}{(D-1)^2(D-2)^2}  \, ,
\nn\\
N^{(D)}_1(\tilde r) =& - \frac{12(D-3)(D-4)}{(D-2)^2} \frac{c_0^2}{\tilde r^{2(D-1)}} + n_1
\end{align}
as the leading order correction.  One obvious difference between this case and the four dimensional case is that the lapse can no longer be simply set equal to unity: in all dimensions greater than four, the  solutions are characterized by two functions, $f(\tilde r)$ and $N(\tilde r)$.

\begin{figure}[htp]
\centering
\includegraphics[width=0.3\textwidth]{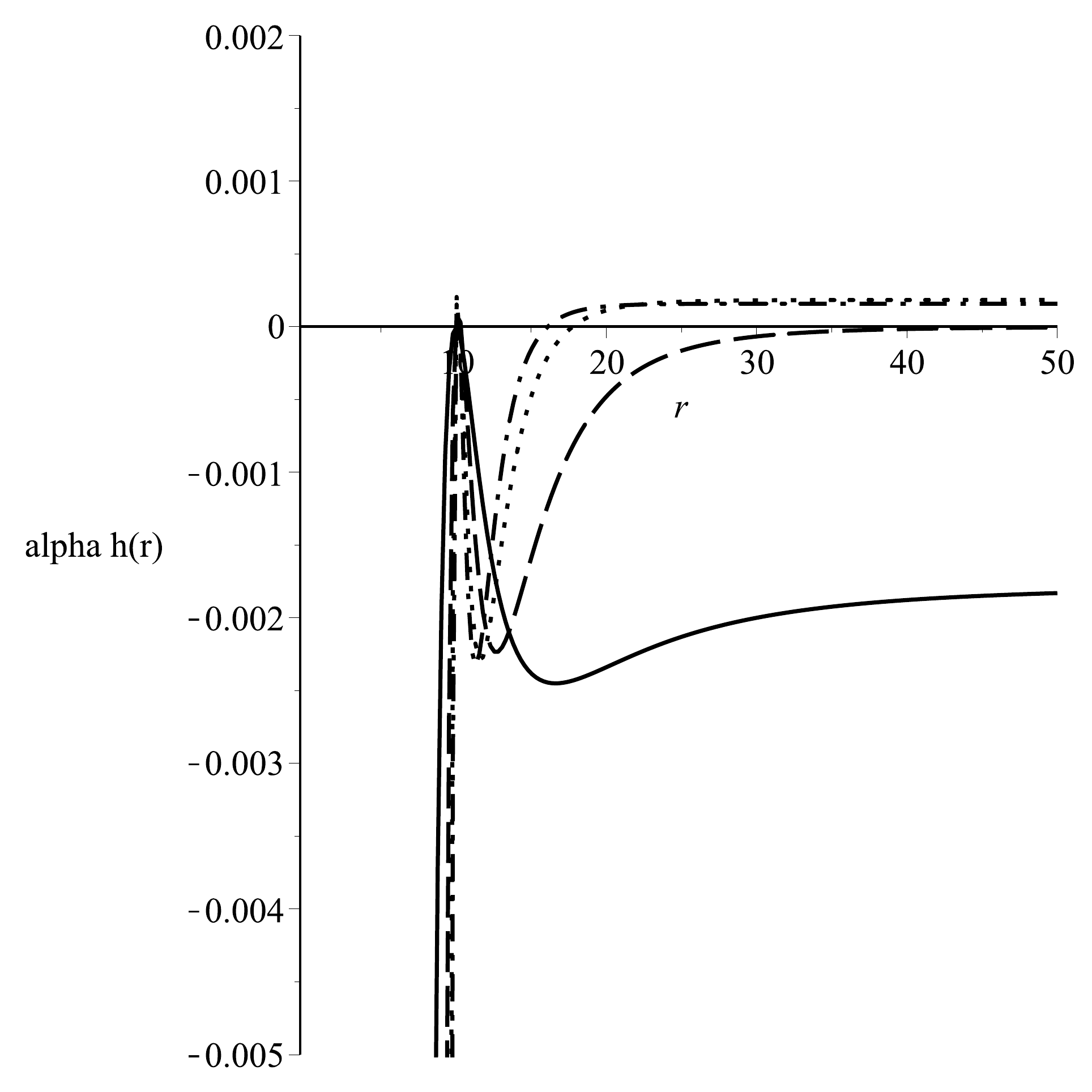}
\includegraphics[width=0.3\textwidth]{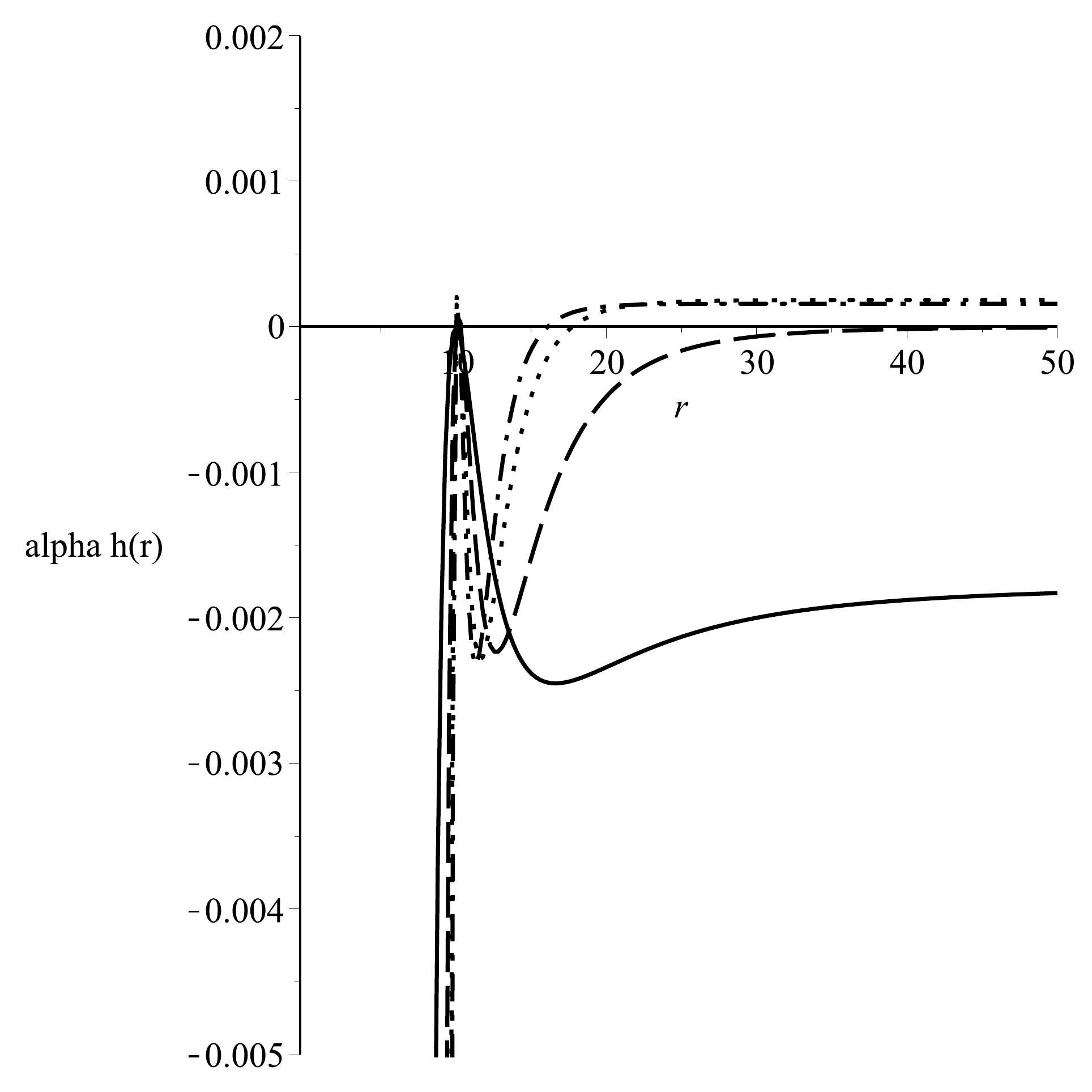}
\includegraphics[width=0.3\textwidth]{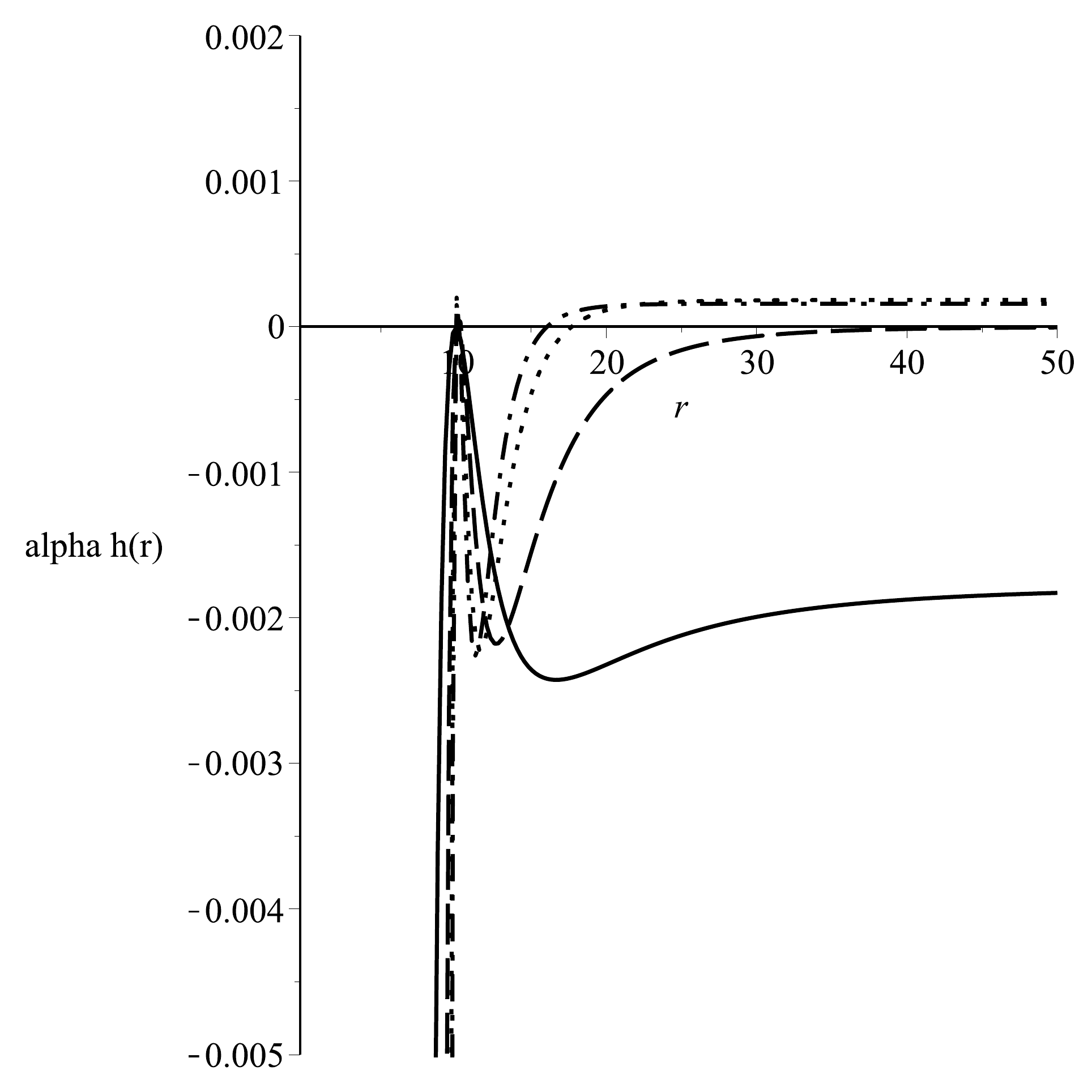}
\\ \quad \quad
\includegraphics[width=0.3\textwidth]{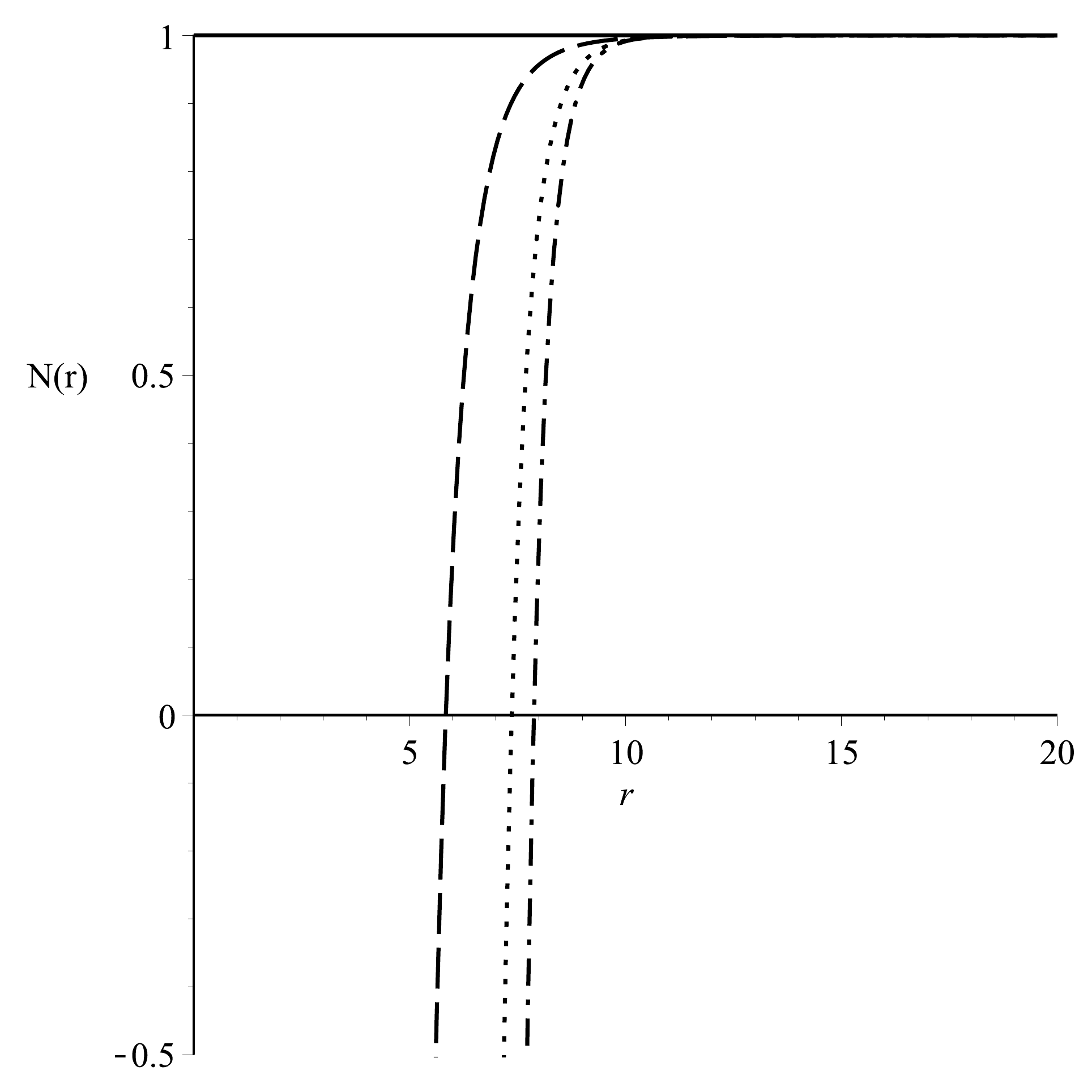}
\includegraphics[width=0.3\textwidth]{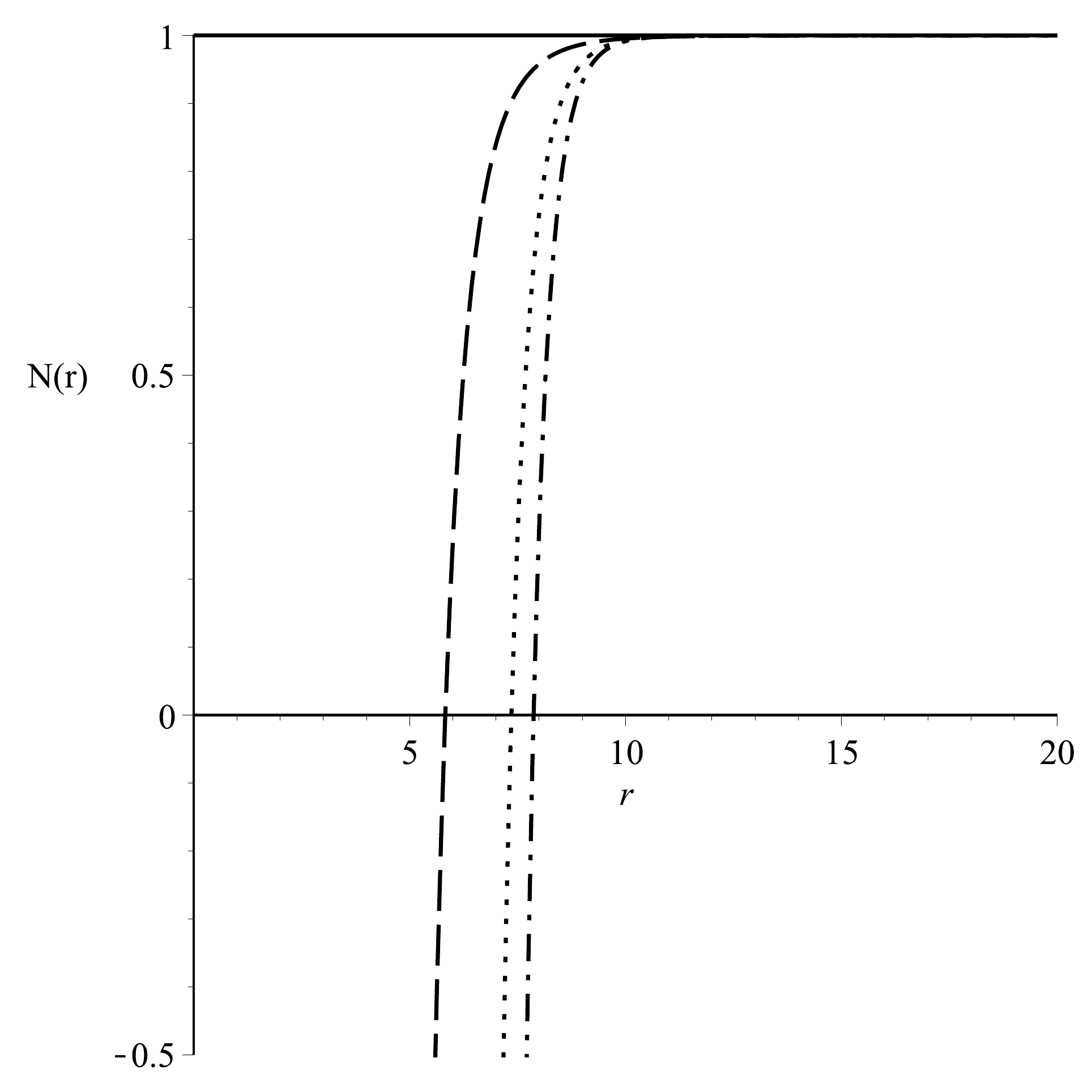}
\includegraphics[width=0.3\textwidth]{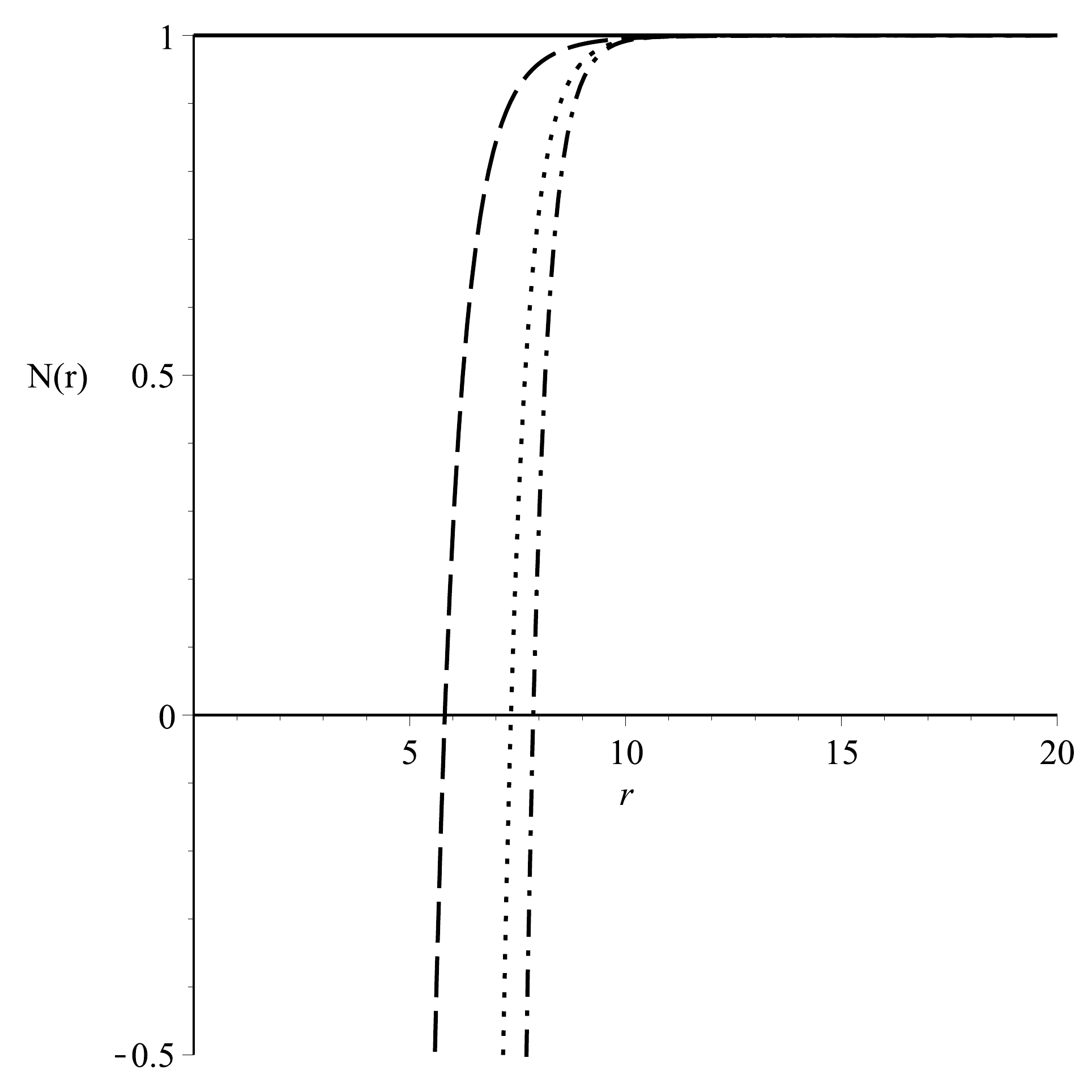}
\\ \quad \quad
\includegraphics[width=0.3\textwidth]{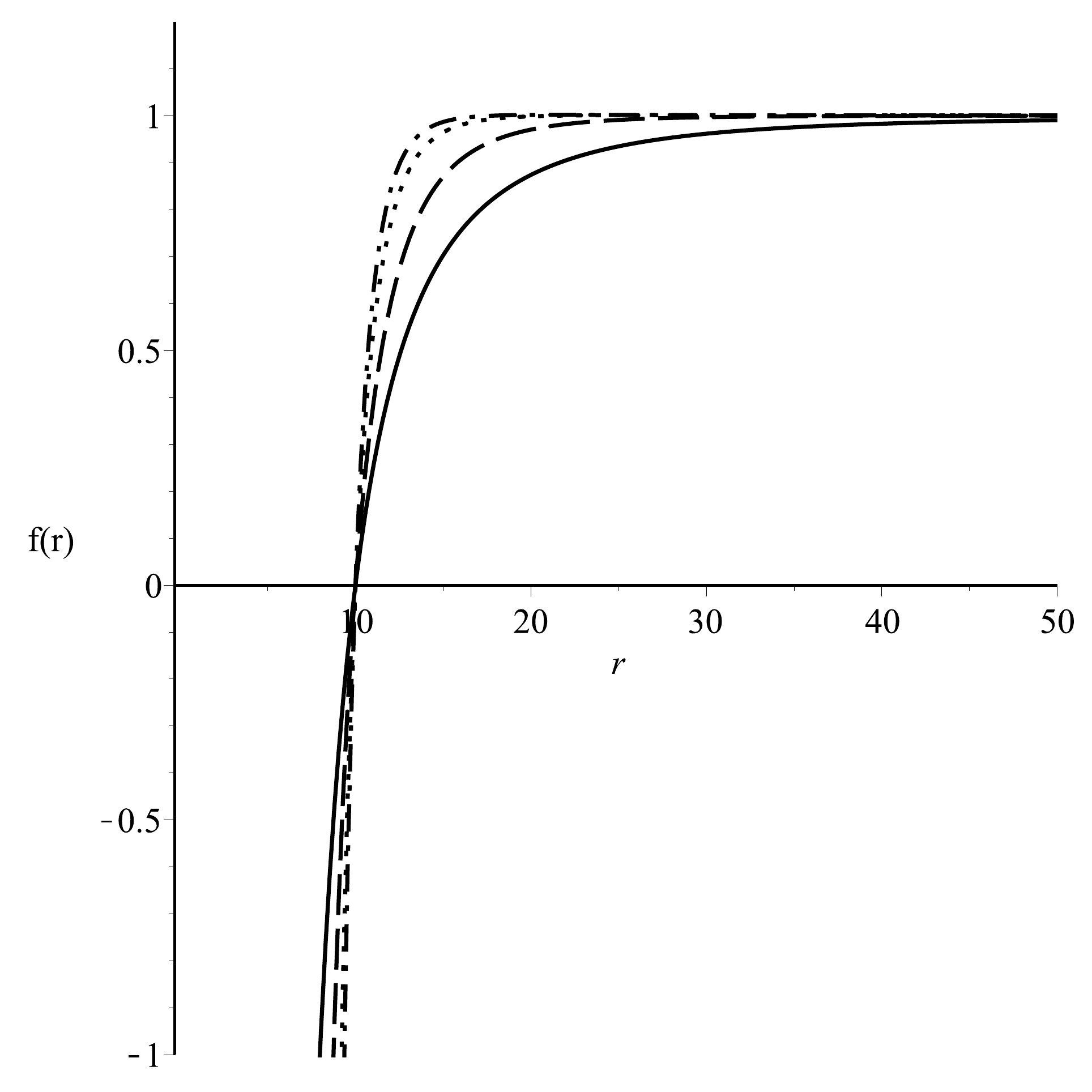}
\includegraphics[width=0.3\textwidth]{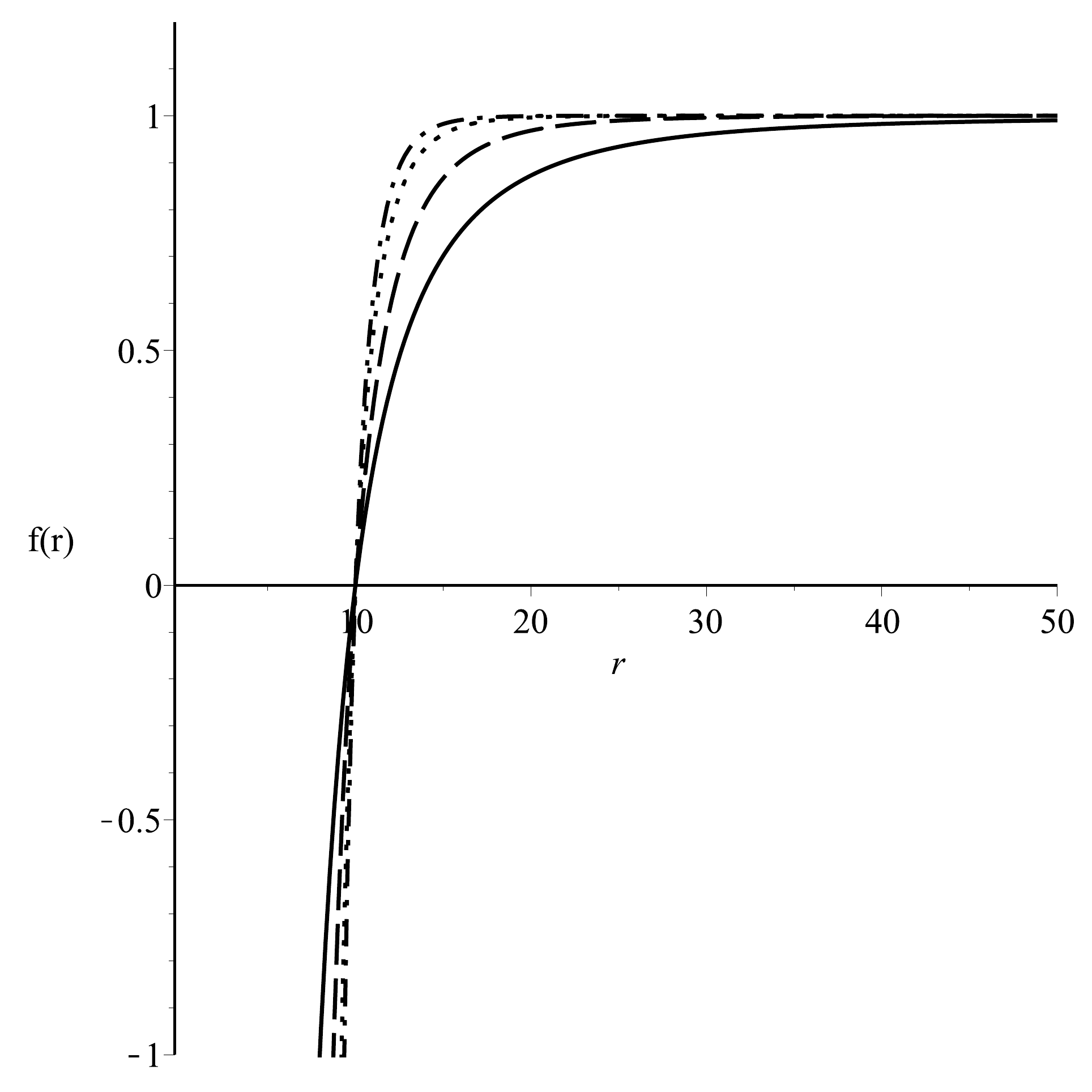}
\includegraphics[width=0.3\textwidth]{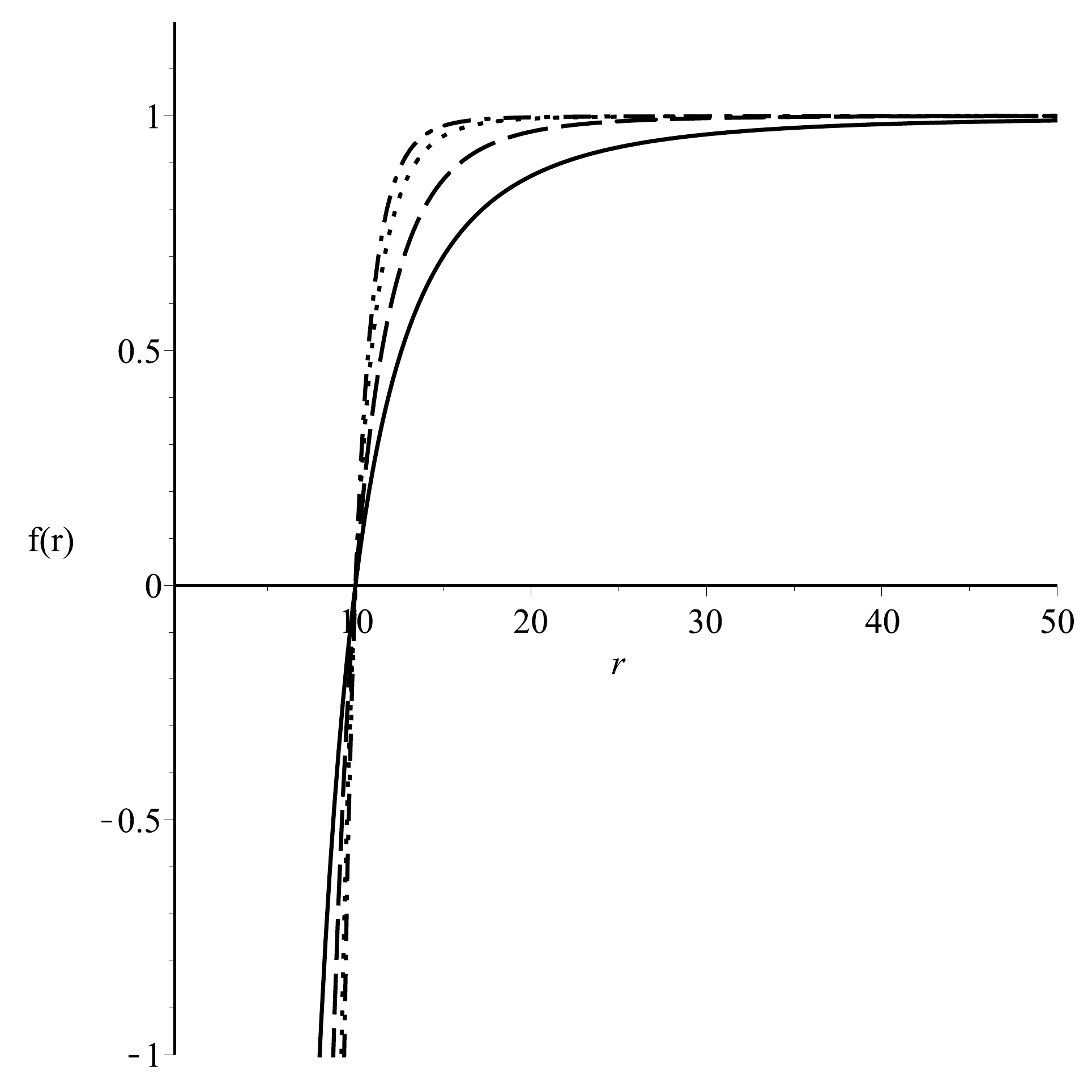}
\caption{{\bf First order corrections in higher dimensions}: {\it Top row}: Plots of the first order corrections, $h_1^{(D)}(\tilde r)$ for spherical, flat and hyperbolic geometries (left, center, right) for a variety of dimensions. {\it Middle row}: Plots of the lapse function, $N(\tilde r)$ with corrections up to ${\cal O}(\alpha)$.  We have set $n_1 = 0$ in these plots. {\it Bottom row}: Plots of the metric function, $f(\tilde r)$ with the inclusion of the first order corrections shown in the top row for $\alpha=0.001$.   In all cases, $c_0$ and $c_1$ were chosen so that the horizon is at $\tilde r_+=10$.  In each plot the solid, dashed, dotted and dash-dotted lines correspond to $D=4, 6, 9, 11$, respectively. Note that, for small enough $\tilde r$ within the horizon, the first order correction terms become comparable to the zeroth order terms, and the termination of the series at first order breaks down. }
\label{fig:higher_d_corrections}
\end{figure}

We have verified that these results hold explicitly up to $D=12$.  In each case the constant $c_1$ can be regarded as an order $\alpha$ correction to the mass of the black hole.  These constants can be fixed via appropriate boundary conditions, e.g. requiring the location of the black hole horizon to be fixed. The constant $n_1$ is an integration constant coming from the differential equations determining the lapse. Higher order corrections are computed quite easily for a given dimension, but determining the general form is a non-trivial process.  We note that both $h_1^{(D)}(\tilde r)$ and $N_1^{(D)}(\tilde r)$ vanish identically in $D=3$ and furthermore that the perturbative correction to the cosmological constant,
\be 
\frac{32(D-3)(D-6)}{(D-1)^2(D-2)^2} \alpha
\ee
vanishes in $D=3$ and $D=6$.  In the latter case, this is because the contribution to the equations of motion of all six dimensional cubic gravities vanish identically for an Einstein space (cf. footnote 41 in \cite{Bueno:2016xff}). That is, when $c_0=0$ in $D=6$, the cubic curvature terms cannot contribute to the equations of motion.  Furthermore, we point out that $N_1^{(D)}(\tilde r)$ vanishes in four dimensions, consistent with the fact that the lapse can be set to unity in four dimensions (amounting to the choice $n_1 = 0$).  In Figure~\ref{fig:higher_d_corrections} we display $h_1^{(D)}(\tilde r)$, $N_1^{(D)}(\tilde r)$,  and $f(\tilde r)$ to ${\cal O}(\alpha)$ for a variety of dimensions.  In the appendix we write explicitly some of the higher order corrections.  

\begin{figure}[htp]
\centering
\includegraphics[width=0.4\textwidth]{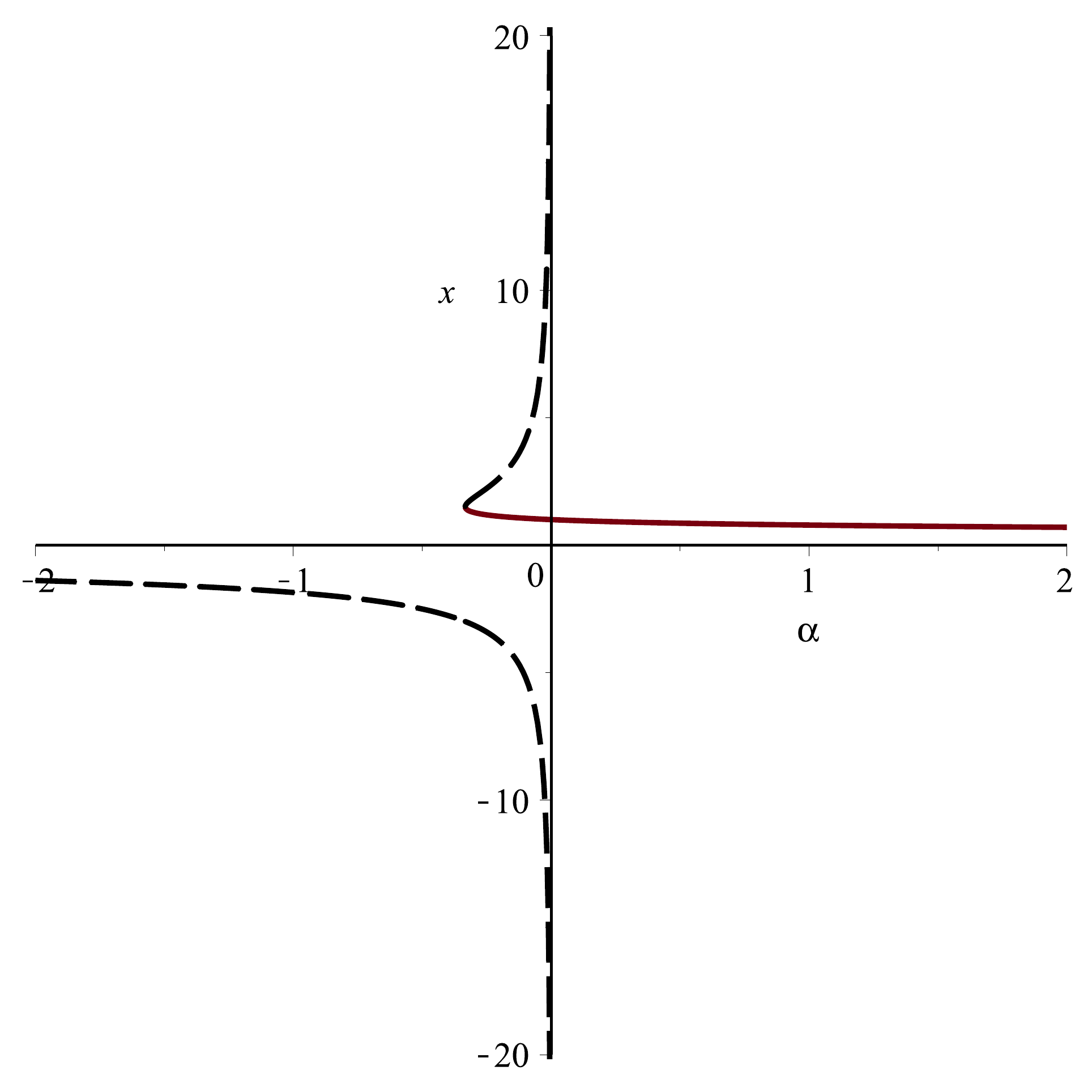}
\includegraphics[width=0.4\textwidth]{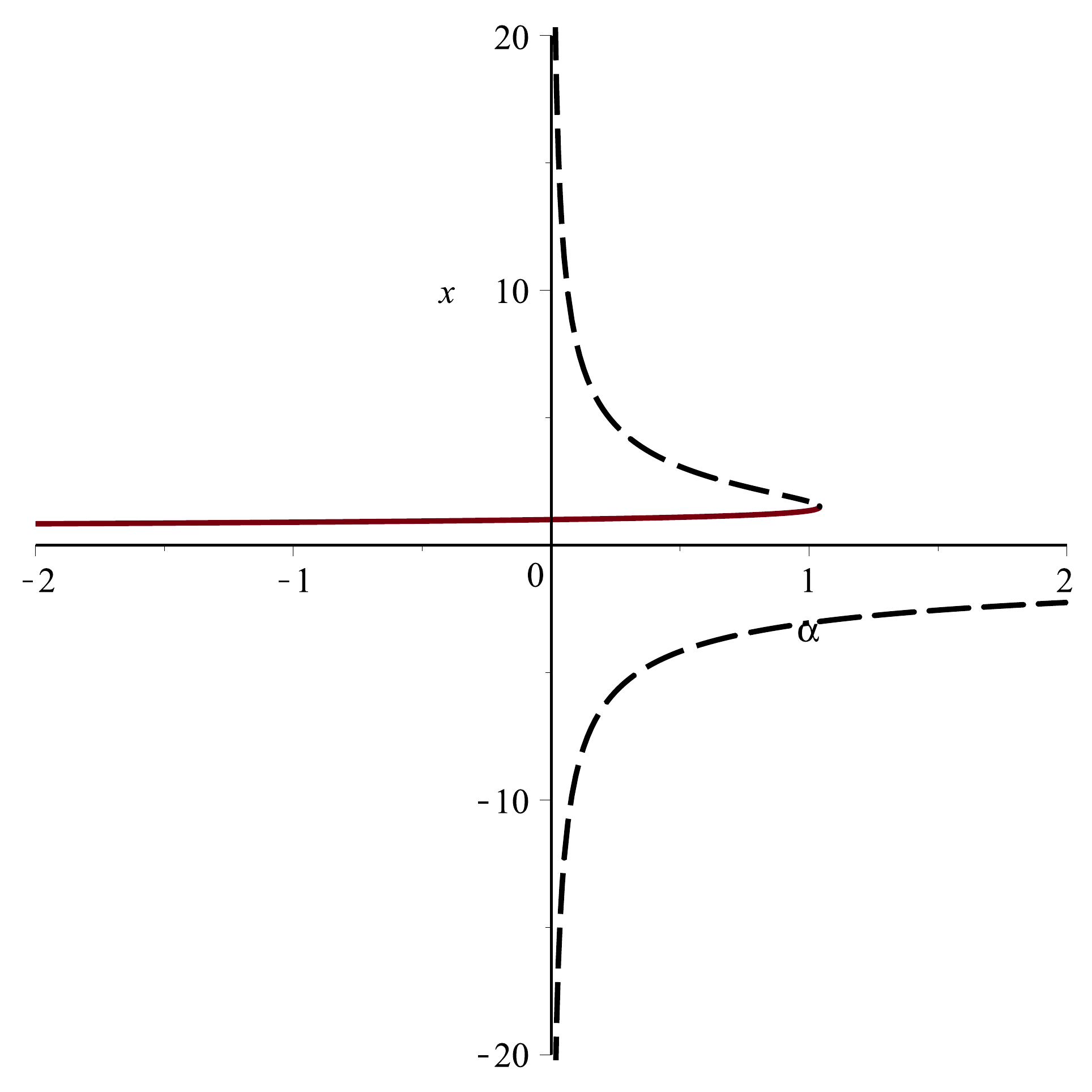}
\caption{{\bf Ghost condition in higher dimensions}: Plots of $x= \Lambda_{\rm eff}/\Lambda$ vs. $\alpha = \kappa^2 \Lambda^2 \lambda$ for five dimensions (left) and seven dimensions (right). Here the black, dashed lines indicate that for these branches the graviton is a ghost, while for the branch indicated by a solid red line is ghost free.  Note the the plots for $ D > 7$ are qualitatively identical to the $D=7$ case: the only ghost free branch is the Einstein branch. }
\label{fig:ghost_higher_d}
\end{figure}

We can go further and analyse the asymptotics of the solutions, incorporating the restrictions placed on the theory by requiring the absence of ghosts.  Considering Eqs.~\eqref{eff_cosmo} and \eqref{ghost_free} for the cases of $D \ge 5$ (but not $D=6$) we find that only the Einstein branch is free from ghosts.  We highlight this graphically in Figure~\ref{fig:ghost_higher_d}.  In the $D=6$ case, ${\cal P}$ does not contribute to the equations of motion for an Einstein metric and the spectrum agrees exactly with Einstein gravity.   

It is useful for thermodynamic analysis  to compute the near horizon solution.  As in the case of four dimensions, we write the near horizon metric functions as
\begin{align} 
f(\tilde r) &= \sum_{i=1}^{i_{\rm max}} A_i (\tilde r-\tilde r_+)^i \, ,
\nn\\
N(\tilde r) &= \sum_{i=0}^{i_{\rm max}} N_i (\tilde r - \tilde r_+)^i
\end{align}
where the $A_i$'s and $N_i$'s are dimensionless coefficients and we require the metric function to vanish linearly as $\tilde r \to \tilde r_+$.  Note that we include a term $N_0$ at this point. Substituting these expressions into the field equations produces a series of complicated relationships which must be satisfied by the $A_i$'s and $N_i$'s.  For example, at order $(\tilde r-\tilde r_+)$, the following relationship must hold:
\begin{align}\label{eqn:nh_o1_higherd} 
0 =& \frac{(D-1)^3(D-2)^2}{432} - \frac{(D-1)^2(D-2)^2 A_1 \tilde r_+}{432} + \frac{(D-1)^2(D-2)^2(D-3) k}{ 432 \tilde r_+^2}
\nn\\
& + \frac{(D-3) \alpha k A_1^2}{9}  + \frac{(D-3) (D-4)\alpha \tilde r_+^3 A_1^3 }{54} - \frac{2(D-3)(D-4)(D-5) \alpha k^3}{27 \tilde r_+^6} 
\nn\\
& + \frac{2 (D-3) \alpha \tilde r_+ A_1^2 (2k + \tilde r_+^3 A_1)}{9} \frac{N_1}{N_0}
\end{align}
where, once again, this expression has been explicitly checked up to $D=12$.  Similar expressions can be written at higher orders, but they rapidly become unwieldy and for this reason we do not present them here, but rather discuss the general character.  The parameter $N_0$ is not fixed at any order, and we therefore take it equal to unity for convenience.  The above expression is linear in $N_1$ but cubic in $A_1$: at second order, the equivalent expression is linear in both $A_2$ and $N_2$.  At higher orders in $(\tilde r-\tilde r_+)$, say at order $n$, $a_{n+1}$ and $N_{n+1}$ will occur and can be determined in terms of two free parameters, which can be taken to be any one of $\{A_1, N_1\}$ plus one of $\{A_2, N_2\}$.  Thus, the black holes of this theory are in general characterized by two free parameters in $D > 4$. 

In general, we see that Eq.~\eqref{eqn:nh_o1_higherd} is cubic in $A_1$, which can be seen as determining the three separate branches of the solution.  However, as we saw earlier, it is only the Einstein branch of this theory which is free from ghosts.  Thus it is natural to investigate what happens when $\alpha$ is small.  To explore this, we write
\begin{align}\label{eqn:small_alpha_ansatz}
A_1 &= \sum_{i=0} \alpha^i A_1^{(i)} \, ,
\quad
A_2 = \sum_{i=0} \alpha^i A_2^{(i)} \, ,
\nn\\
N_1 &=\sum_{i=0} \alpha^i N_1^{(i)} \, ,
\quad
N_2 =\sum_{i=0} \alpha^i N_2^{(i)} \, ,
\end{align}
and so on.  Substituting these expressions into the field equations it is found that the free parameters are fixed order by order in $\alpha$.  To illustrate this, we shall write a few of the more useful expressions:
\begin{align}\label{eqn:some_first_order_expansions}
N_1 &= \frac{24\alpha (D-1)(D-3)(D-4) (\tilde r_+^2 + k)^2}{(D-2)^2 \tilde r_+^5} + {\cal O}(\alpha^2) \, ,
\nn\\
N_2 &= -\frac{12\alpha (D-1)(D-3)(D-4)(2D-1) (\tilde r_+^2 + k)^2 }{(D-2)^2 \tilde r_+^6} + {\cal O}(\alpha^2) \, ,
\nn\\
A_1 &= \frac{(D-1)\rt^2 + (D-3)k}{\rt ^3} + \frac{\alpha (D-3)}{(D-2)^2} \bigg( \frac{8 (D-1)(D-4)}{ \rt} 
+ \frac{24k(D^2 - 7D + 14)}{ \rt^3} 
\nn\\
&+ \frac{24 k^2 (D-3) (D^2 - 7d + 16) }{(D-1) \rt ^5 } + \frac{8k^3(D^3 - 12 D^2 + 53 D - 82)}{(D-1) \rt^7} \bigg) 
\nn\\
&+ \frac{\alpha^2 (D-3)^2}{(D-2)^4} \bigg( \frac{192(D-1)(D-4)(13D-16)}{\rt} 
+ \frac{192 k (D-4)(65D^2 - 203 D + 178)}{\rt^3} 
\nn\\
&+ \frac{384 k^2 (65 D^4 - 586 D^3 + 1885 D^2 - 2744 D + 1692)}{(D-1) \rt^5 } 
\nn\\
&+ \frac{384 k^3 (65 D^5 - 709 D^4 + 3003 D^3 - 6481 D^2 + 7694 D - 4292)}{(D-1)^2 \rt^7} 
\nn\\
&+ \frac{192 k^4 (65 D^6 - 832 D^5 + 4342 D^4  - 12284 D^3 + 21097 D^2 - 21980 D + 10792)}{(D-1)^3 \rt^9} 
\nn\\
&+ \frac{192 k^5 (D-3)(13 D^5 - 139 D^4 + 585 D^3 - 1341 D^2 + 1938 D - 1456)}{(D-1)^3 \rt^{11}} \bigg) + {\cal O}(\alpha^3) \, ,
\nn\\
A_2 &= \frac{6k - (\rt^2 + k)D(D-1) }{2 \rt^4} - \frac{\alpha (D-3) }{(D-2)^2 \rt^8} \bigg(4 \rt^6 (D-1)(10D^2 - 37D + 24) 
\nn\\
&+  12k \rt^4 (10D^3 - 57 D^2 + 111D - 80) + \frac{12 k^2 \rt^2 (10D^4 - 77 D^3 + 232 D^2 - 321 D + 136)}{(D-1)} 
\nn\\
&+ \frac{4k^3 (10 D^4 - 87 D^3 + 296 D^2 - 451 D + 192)}{(D-1)} \bigg) + {\cal O}(\alpha^2) \, .
\end{align}

These expressions can be evaluated to higher order in $\alpha$, but the conclusion remains the same: the freedom to choose two parameters is lost when one requires that the solution has a smooth $\alpha \to 0$ limit.  The reason for this can be easily seen by performing small $\alpha$ series expansions of the equations.  For example, rearranging Eq.~\eqref{eqn:nh_o1_higherd} for $N_1$ and expanding near $\alpha=0$, while setting $A_1$ to be that given in Eq.~\eqref{eqn:small_alpha_ansatz} one finds that (ignoring proportionality constants) it goes like 
\be 
N_1 \sim \frac{\rt^3 A_1^{(0)} + 3k + \rt^2 - D(\rt^2 +k)}{\alpha} +  {\cal O}(\alpha^0) \, .
\ee
Thus, avoiding a divergence as $\alpha \to 0$ in this term requires fixing $A_1^{(0)}$ to be that given in Eq.~\eqref{eqn:some_first_order_expansions}.  Similar problems occur at all orders and completely use up the free parameters.  This is not unlike the four dimensional case where we saw a similar result in Section~\ref{sec:4d_case_expansions}.  Since we wish to study the ghost free (i.e. Einstein) branch, we will then work with the expansions given above in Eq.~\eqref{eqn:some_first_order_expansions} in what follows.

\begin{figure}
\centering
\includegraphics[width=0.4\textwidth]{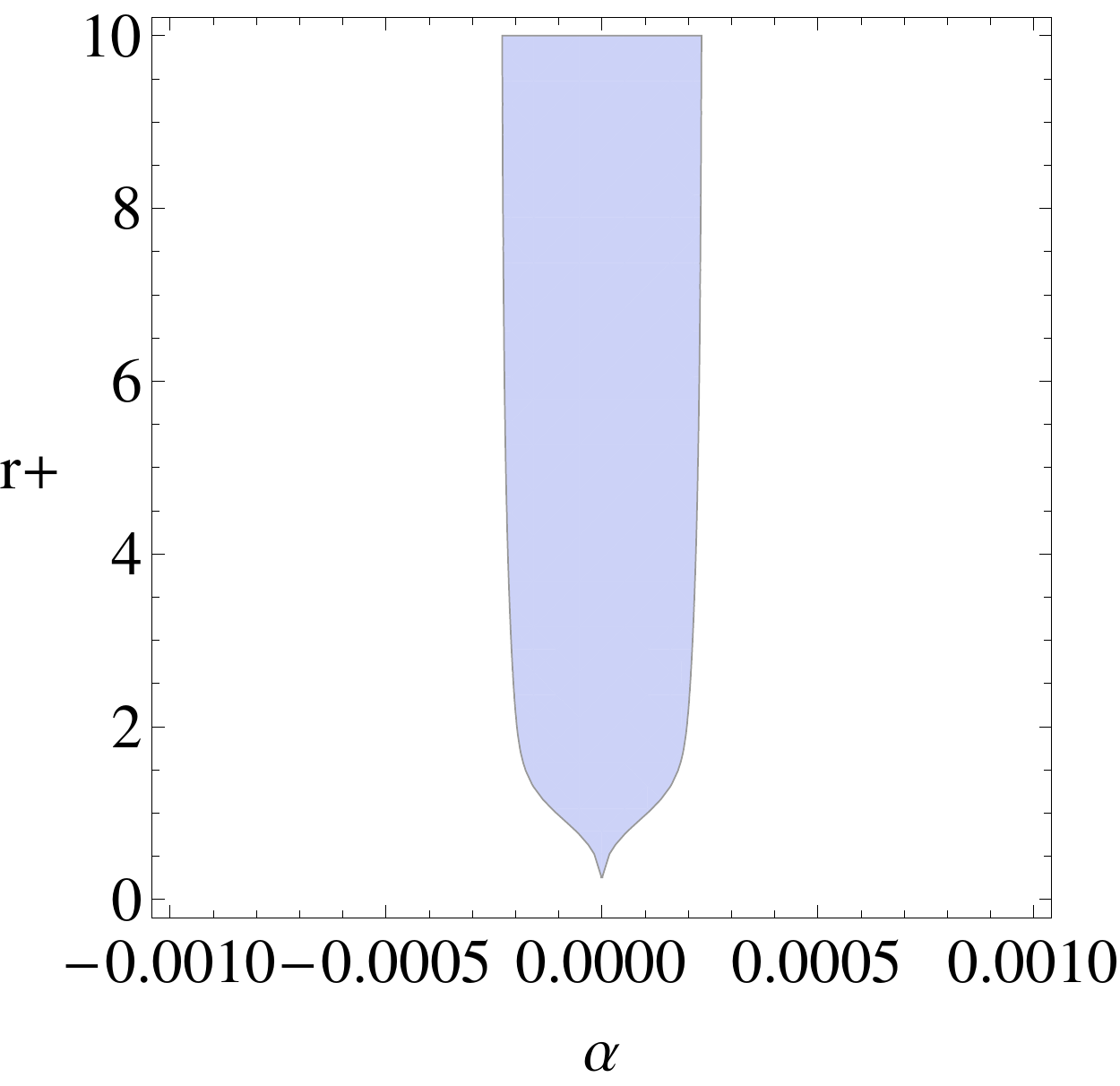}
\includegraphics[width=0.4\textwidth]{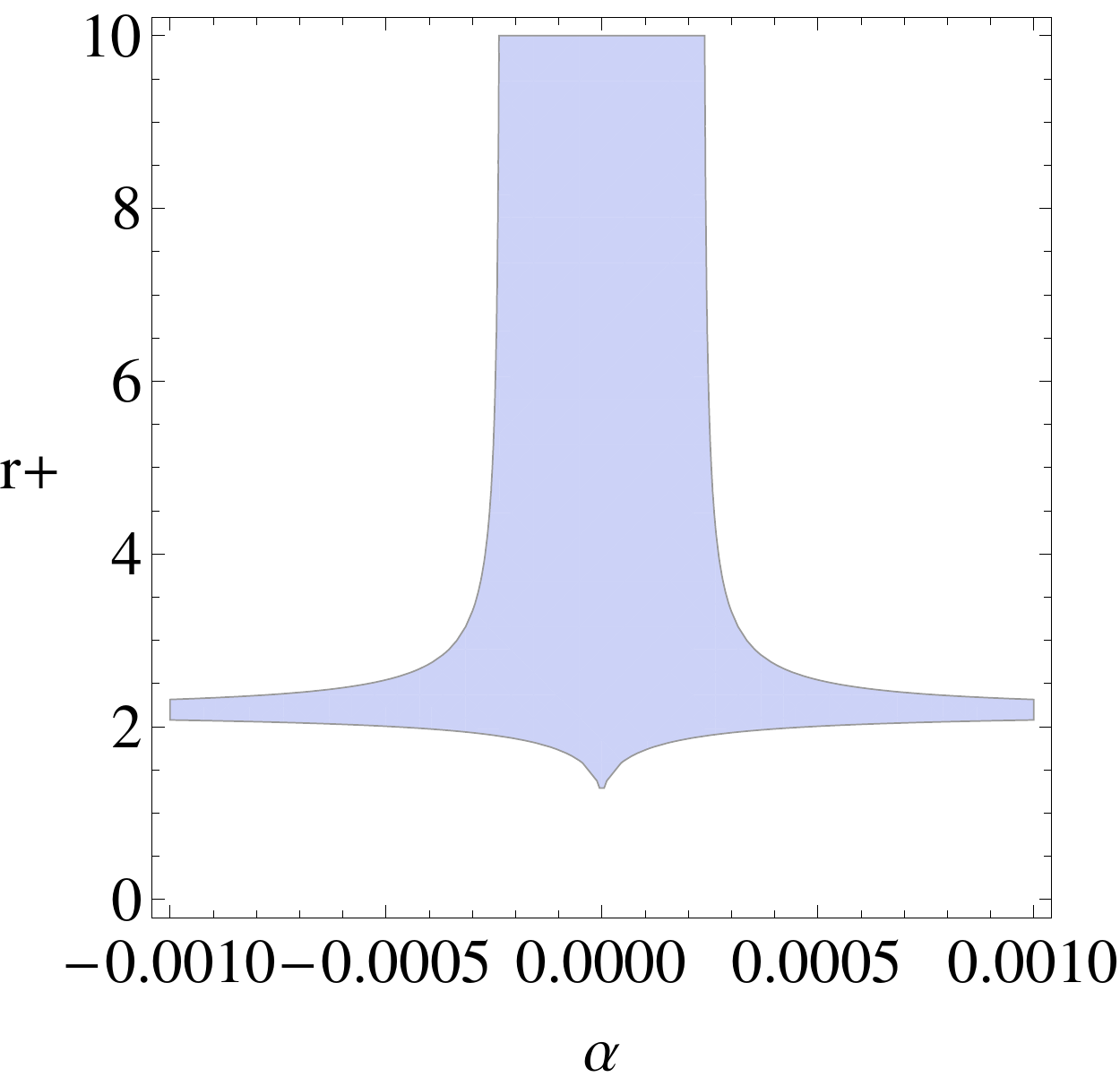}
\caption{{\bf Error profile for $D=5$}: Here the left plot corresponds to $k=+1$ and the right plot to $k=-1$.  In each case, for parameters within the blue shaded region we have $A_1 > 0$ and $\varepsilon |\alpha^2 A_1^{(2)}| > |\alpha^3 A_1^{(3)}|$ with $\varepsilon$ chosen as $1/10$, justifying terminating the series at order $\alpha^2$.   }
\label{fig:error_5d}
\end{figure}

To begin our consideration of the thermodynamics, we compute the Iyer-Wald entropy~\cite{Iyer:1994ys}.  We find that for the metric Eq.~\eqref{higherDmetric}, the entropy takes the form
\begin{align} 
S &= \frac{2 \pi A}{\kappa} \left[1 + \frac{24(D-3) A_1 \alpha}{(D-2)(D-1)^2 \tilde r_+}\left(4 k + \tilde r_+^3 A_1 \right)   \right]\, ,
\end{align}
where, in the above expression  the area is given by 
\be 
A = |N_0| \omega_{(k) D-2} (\ell \tilde r_+)^{D-2}=\omega_{(k) D-2} r_+^{D-2}  
\ee
where $\omega_{(k)d-2}$ is the area of the surface defined by $d\Sigma^2_{(k)d-2}$ and we have chosen to set $N_0=1$. The temperature can be computed by requiring the absence of conical singularities in the Euclidean sector
\be 
T = \frac{\tilde r_+^2 |N(\rt)| f'(\tilde r_+)}{4 \pi \ell } = \frac{\tilde r_+^2 A_1}{4 \pi \ell} \, .
\ee
 Remarkably, the higher curvature corrections to the entropy are completely characterized in terms of $A_1$.  However, unfortunately, we do not have access to an exact expression for $A_1$ as we did in the four dimensional case.  Therefore, we can only explore the thermodynamics here at a perturbative level.  However, it must be kept in mind that in order for the perturbative results to be valid, the higher order terms must die off rather than grow.  Essentially the effect of this is to introduce a relationship between $\alpha$ and $\rt$ which means that the perturbative results will not be valid for all combinations of $\rt$ and $\alpha$---similar to what was discussed for the four-dimensional case in Section~\ref{sec:4d_limits}. We furthermore require that $A_1 > 0$, so that it describes a black hole rather than a (possibly) singular cosmological solution. In going forward with the analysis, we will work to ${\cal O}(\alpha^2)$ and numerically enforce bounds on the maximum relative error of the ${\cal O}(\alpha^3)$ term, as we highlight in Figure~\ref{fig:error_5d} for $D=5$.  We will focus our attention to the cases of five and six dimensions.
 
\begin{figure}
\centering
\includegraphics[width=0.4\textwidth]{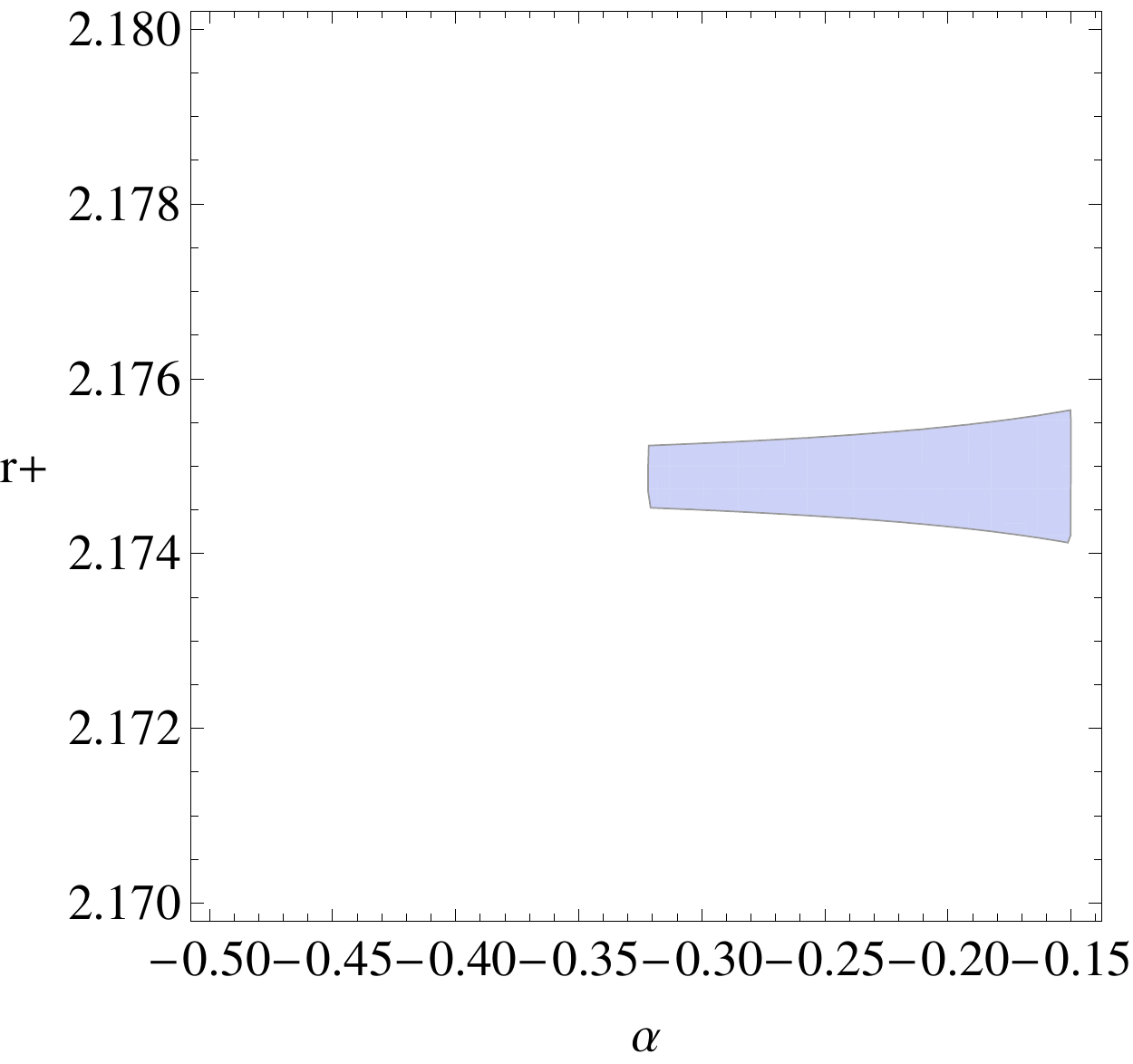}
\caption{{\bf Region of negative entropy}: This plot shows the $D=5$ case with $k=-1$.  The blue shaded region corresponds to negative entropy.  In the $D=6$ case, the negative entropy region corresponds to $A_1 < 0$ and we therefore exclude it. }
\label{fig:negative_entropy}
\end{figure}
   
 Our first thermodynamic consideration is under what circumstances the entropy is positive.  In both five and six dimensions, and for both $k=+1$ and $k=-1$, the entropy is positive essentially for the entire viable parameter space: plotting this parameter space results in plots visually akin to Figure~\ref{fig:error_5d}.  However, the entropy can be negative for very small portions of the viable parameter space when $k=-1$ for $D=5$, an example is shown in Figure~\ref{fig:negative_entropy}.  In the case where $D=6$ and $k=-1$, the regions of negative entropy correspond to $A_1 < 0$ and we therefore exclude them on other grounds.

The natural next step is to investigate the first law of thermodynamics for these black holes. To obtain an expression for the thermodynamic mass we can integrate the first law of thermodynamics.  In doing so we must remain aware that we are working with perturbative solutions, and therefore they are not valid for arbitrary combinations of $\alpha$ and $\rt$ (see Figure~\ref{fig:error_5d}).  We begin by constructing the first law of thermodynamics in $D=5$ and $D=6$.  Since we are working in dimensionless units the standard extended first law,
\be 
dM = TdS + V dP + \Psi d\lambda \, ,
\ee
is modified according to
\be 
\lambda = \frac{\alpha}{\kappa^4 P^2} \Rightarrow d \lambda = \frac{d\alpha}{\kappa^4 P^2} - \frac{2 \alpha dP}{\kappa^4 P^3} 
\ee
giving
\be 
dM = TdS + \left(V - \frac{2 \alpha}{\kappa^4 P^3}  \right)dP + \frac{\Psi}{\kappa^4 P^2} d\alpha \, .
\ee
where as before we have identified,
\be 
P = -\frac{\Lambda}{\kappa} \, .
\ee
Working first in $D=5$, and to ${\cal O}(\alpha^2)$, we find that the following expressions for the mass and volume satisfy the first law and Smarr formula: 
\begin{align}
M_{D=5} =&  \frac{\omega_{(k) 3}\ell^2}{\kappa} \left[ \frac{3}{2} \tilde r_+^2 (k + \tilde r_+^2) + \frac{2 \alpha}{ 3 \tilde r_+^2} \left(40 \rt^6 + 48 k^2 \rt^2 \ln (\rt) + 108 k \rt^4 + 5k^3  \right)  \right. 
\nn\\
&\left. + \frac{32 \alpha^2}{27 \rt^6} \left(696 \rt^{10} + 15096 k^2 \rt^6 \ln (\rt) + 5256 k \rt^8 - 5214 k^3 \rt^4 - 873 k^4 \rt^2 - 76 k^5   \right)   \right]\, , 
\nn\\
V_{D=5} =& \omega_{(k) 3}\ell^4 \bigg[ \frac{\tilde r_+^4}{4}   + \frac{8 \alpha}{3} ( \rt^4 + 2 k^2 \ln (\rt) + 3 k \rt^2 - k^2 )
\nn\\
& + \frac{16 \alpha^2}{27 \rt^4} ( 1000 \rt^8 + 15096 k^2 \rt^4 \ln (\rt) + 6432 k \rt^6 + 1820 k^2 \rt^4 - 3820 k^3 \rt^2 - 351 k^4
) \bigg] \, ,
\nn\\
\Psi_{D=5} =& \frac{ \omega_{(k) 3} \kappa}{ \ell^2 \rt^2} \bigg[ 24( -8 \rt^6 + 48 k^2 \rt^2 \ln (\rt) - 12 k \rt^4 - 84 k^2 \rt^2 - 13 k^3)
\nn\\
&+ \frac{128 \alpha}{3 \rt ^4} \left(1152 \rt^{10} + 30192 k^2 \rt^6 \ln (\rt) + 9408 k \rt^8 - 1824 k^2 \rt^6 \right. 
\nn\\
&\left. - 11808 k^3 \rt^4 - 2235 k^4 \rt^2 - 218 k^5 \right) 
\nn\\
&- \frac{512 \alpha^2}{9 \rt^8} \left((2\rt^2 + k)^3 (1208 \rt^8 + 5492 k \rt^6 + 9234 k^2 \rt^4 + 6799 k^3 \rt^2 + 1846 k^4  \right) \bigg]
\end{align}
and in $D=6$:
\begin{align}
M_{D=6} =& \frac{\omega_{(k)4}\ell^3}{\kappa} \bigg[2\tilde r_+^3 ( k + \tilde r_+^2)  + \frac{6 \alpha}{\rt} (7 \rt^6 + 20 \rt^4 k + 27 \rt^2 k^2 - 2 k^3) 
\nn\\
& + \frac{18 \alpha^2}{25 \rt^5} (2775 \rt^{10} + 18250 k \rt^8 + 85320 k^2 \rt^6 - 65670 k^3 \rt^4 - 8335 k^4 \rt^2 - 756 k^5) \bigg] \, ,
\nn\\
V_{D=6} =& \omega_{(k) 4}\ell^5 \bigg[ \frac{\tilde r_+^5}{5} + \frac{6 \alpha \rt }{5 } \left( 3 \rt^4 + 8 k \rt^2 + 9 k^2 \right) 
\nn\\
& + \frac{36 \alpha^2}{25 \tilde r_+^3} \left( 600 \rt^8 + 3350 k \rt^6 + 12447 k^2 \rt^4 - 6744 k^3 \rt^2 - 449 k^4 \right) \, ,
\nn\\
\Psi_{D=6} =& \frac{ \omega_{(k) 4} \kappa}{ \ell \rt} \bigg[
-12 (25 \rt^6 - 25 k \rt^4 - 585 k^2 \rt^2 + 289 k^3 ) 
\nn\\
&+ \frac{36 \alpha}{\rt^4} \left(9225 \rt^{10} + 65350 k \rt^8 + 329130 k^2 \rt^6 - 272016 k^3 \rt^4 - 36787 k^4 \rt^2 - 3510 k^5 \right) 
\nn\\
&- \frac{324 \alpha^2}{5 \rt^8} \left((\rt^2 + k)^3 (60625 \rt^8 + 192625 k\rt^6 + 215175 k^2 \rt^4 + 102475 k^3 \rt^2 + 17788 k^4  \right)
\bigg]
\end{align}
where the mass has been obtained by integrating the first law of thermodynamics.  Note that the logarithms appearing in the five-dimensional quantities are an artefact of the expansion, which cannot be trusted all the way to $\tilde r_+ = 0$.  It is straightforward to compute these quantities to arbitrary order in any dimension.  However, the resulting expressions are not insightful.  

\begin{figure}[htp]
\centering
\includegraphics[width=0.4\textwidth]{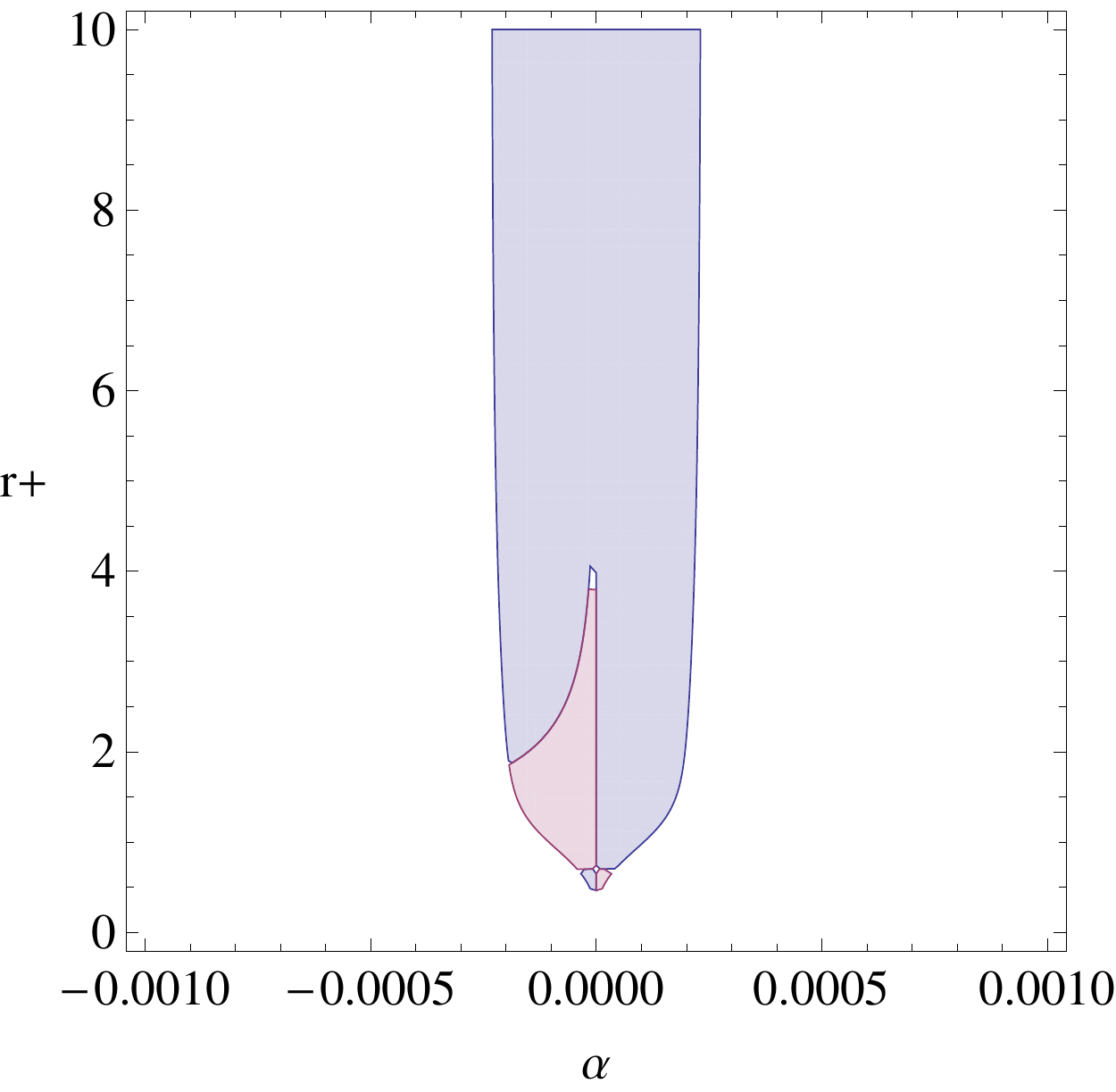}
\includegraphics[width=0.4\textwidth]{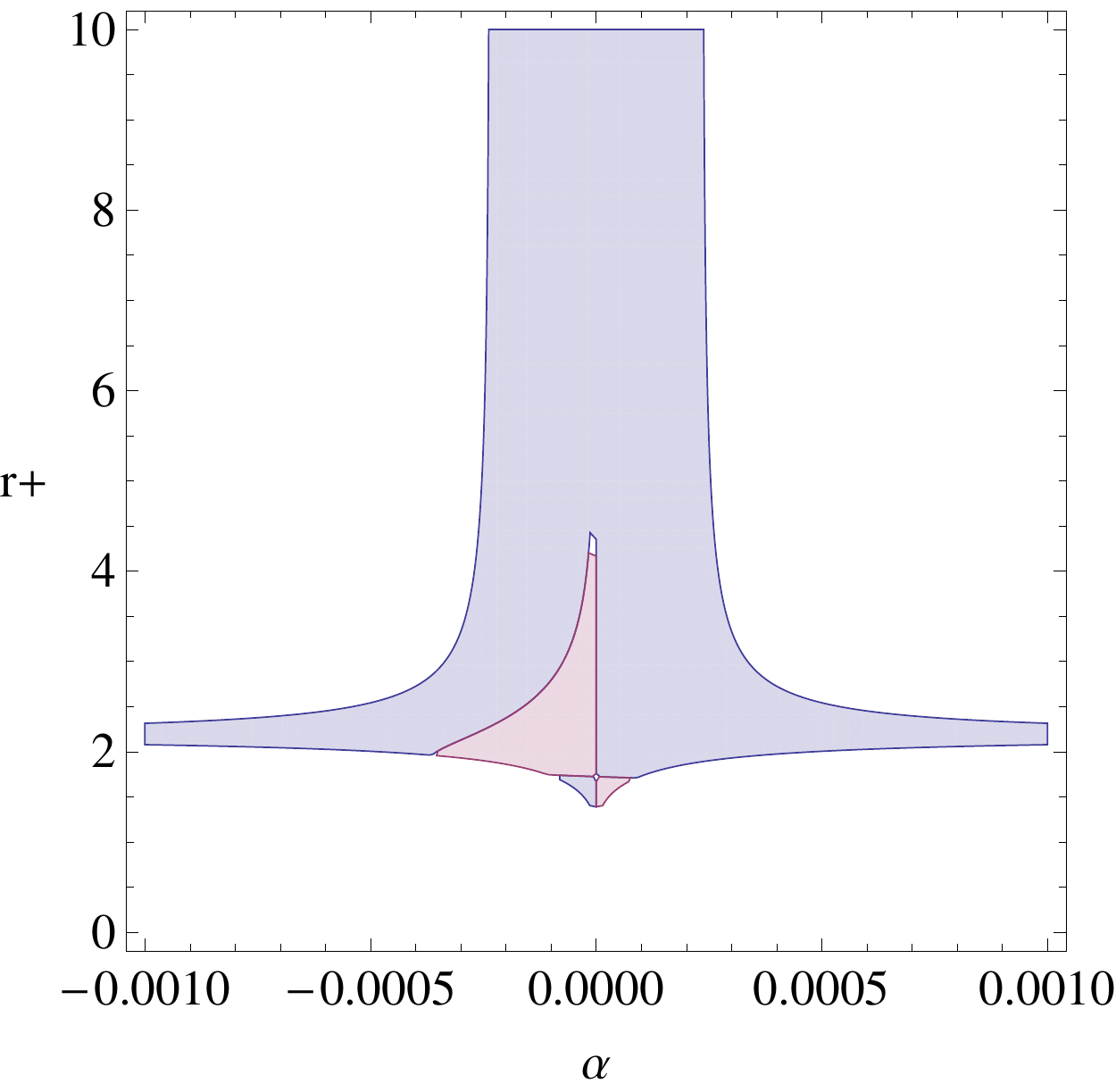}
\includegraphics[width=0.4\textwidth]{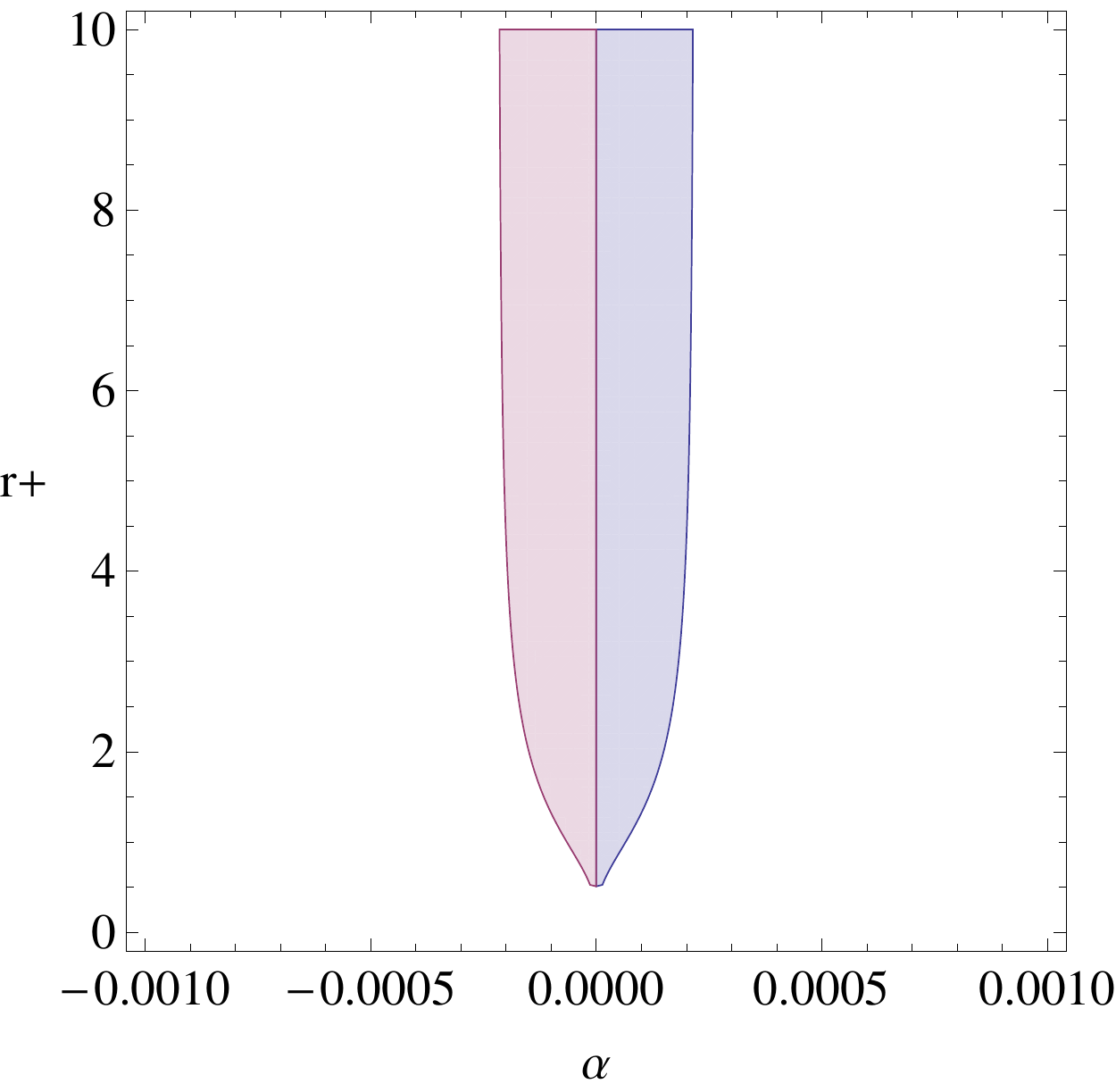}
\includegraphics[width=0.4\textwidth]{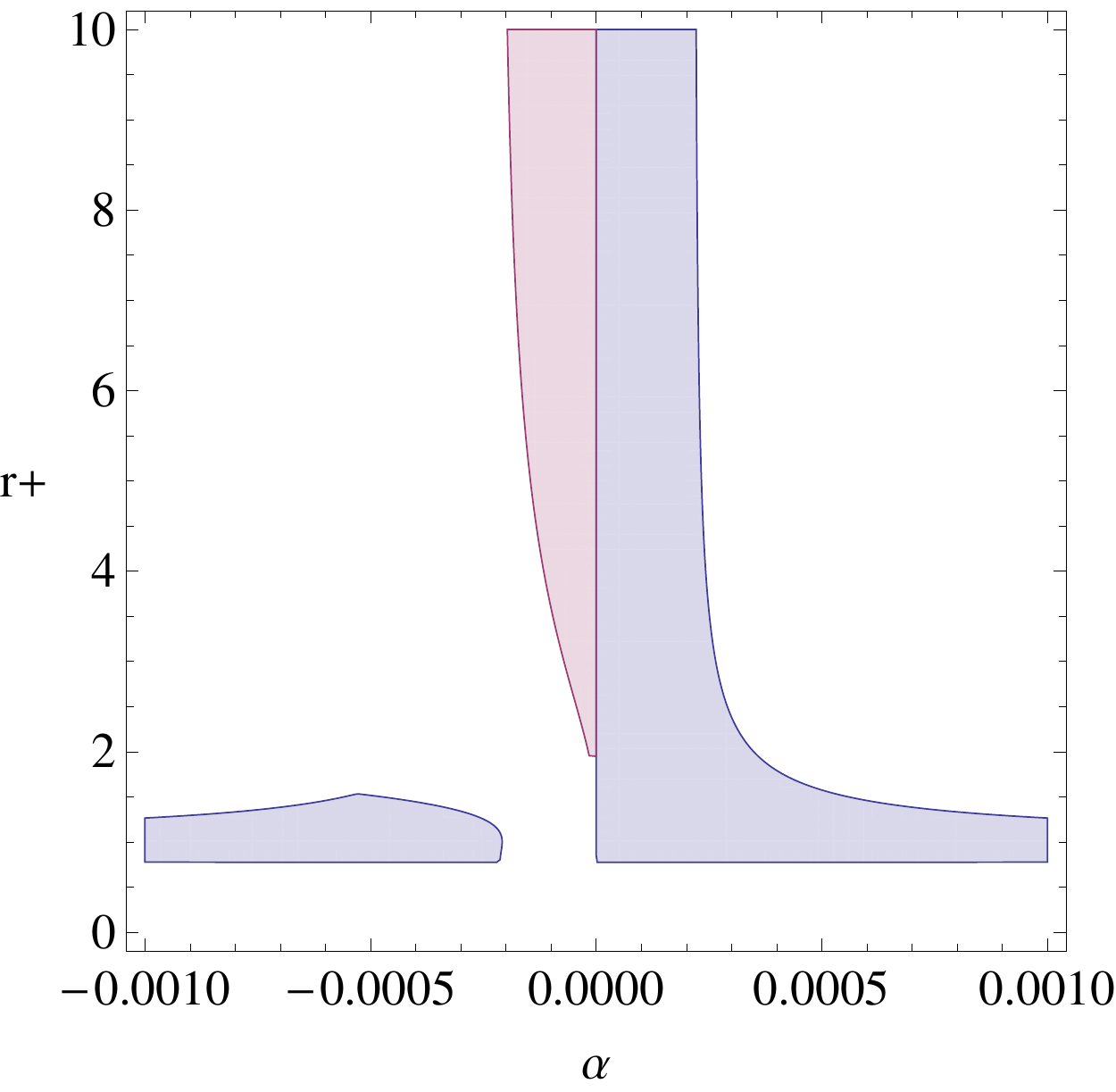}
\caption{{\bf Reverse isoperimetric inequality}: Plots showing regions of parameter space where the reverse isoperimetric inequality \eqref{RIPI} is and is not obeyed.  The plots are for $D=5$ with $k=+1$ and $k=-1$ (top left and right, respectively) and $D=6$ with $k=+1$ and $k=-1$ (bottom left and right, respectively).  In each case, the reverse isoperimetric inequality is obeyed in the blue shaded regions and is violated in the red shaded regions.  In the production of these plots  we have enforced $A_1 > 0$ and $\varepsilon |\alpha^2 A_1^{(2)}| > |\alpha^3 A_1^{(3)}|$ with $\varepsilon$ chosen as $1/10$.   }
\label{fig:iso_ratio}
\end{figure}

It is interesting to note that the thermodynamic volume is not simply the naive geometric volume, but possesses  corrections perturbative in $\alpha$, in contrast to the four-dimensional case discussed in the first part of this paper. This has the effect that, for some parameters, these black holes are super-entropic~\cite{Hennigar:2014cfa}.  That is, their thermodynamic volume does not satisfy the following condition:
\be\label{RIPI}
{\cal R} = \left(\frac{(D-1)V}{\omega_{(k) D-2}} \right)^{\frac{1}{D-1}} \left( \frac{\omega_{(k) D-2}}{A} \right)^{\frac{1}{D-2}} \ge 1 
\ee
which was conjectured for Einstein gravity first in \cite{Cvetic:2010jb} and then later revised to exclude  non-compact horizons in \cite{Hennigar:2014cfa}. Examples of black holes in higher curvature theories of gravity that  violate the reverse isoperimetric inequality were first found in \cite{Brenna:2015pqa} in the study of asymptotically Lifshitz spacetimes, but this is the first example of higher curvature black holes asymptotic to AdS space that violate the inequality.  Note that in the Einstein gravity limit $\alpha \to 0$, the inequality is  an equality, consistent with the conjecture.  Figure~\ref{fig:iso_ratio} shows the regions of parameter space for which the reverse isoperimetric inequality holds when the error in the series expansions used are less than 10\%.  We notice that, provided $\alpha > 0$, the inequality is obeyed in both and $D=6$ when the volume is positive, in $D=5$ the inequality can be violated for both positive and negative $\alpha$.

\begin{figure}[htp]
\centering
\includegraphics[width=0.4\textwidth]{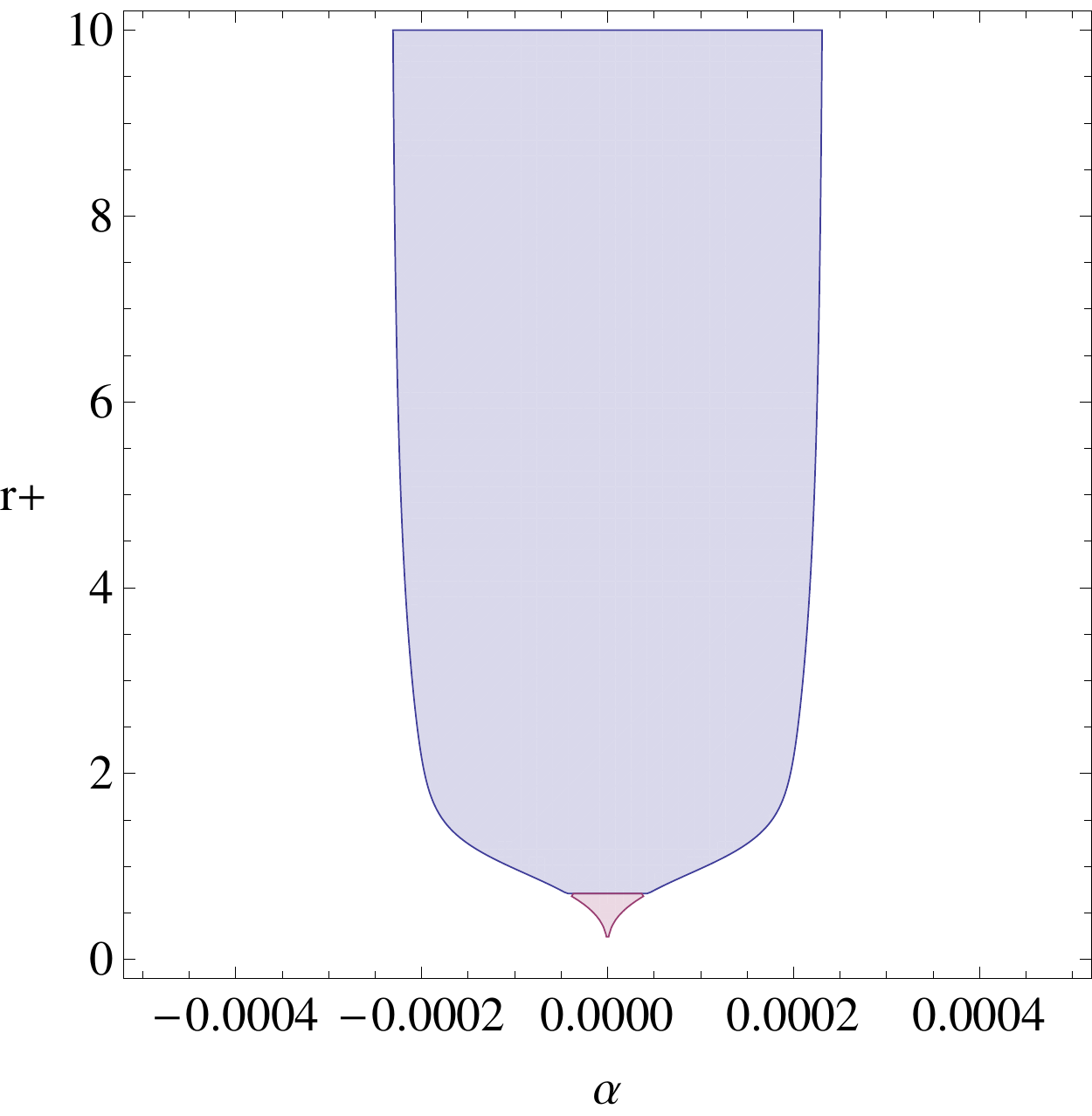}
\includegraphics[width=0.4\textwidth]{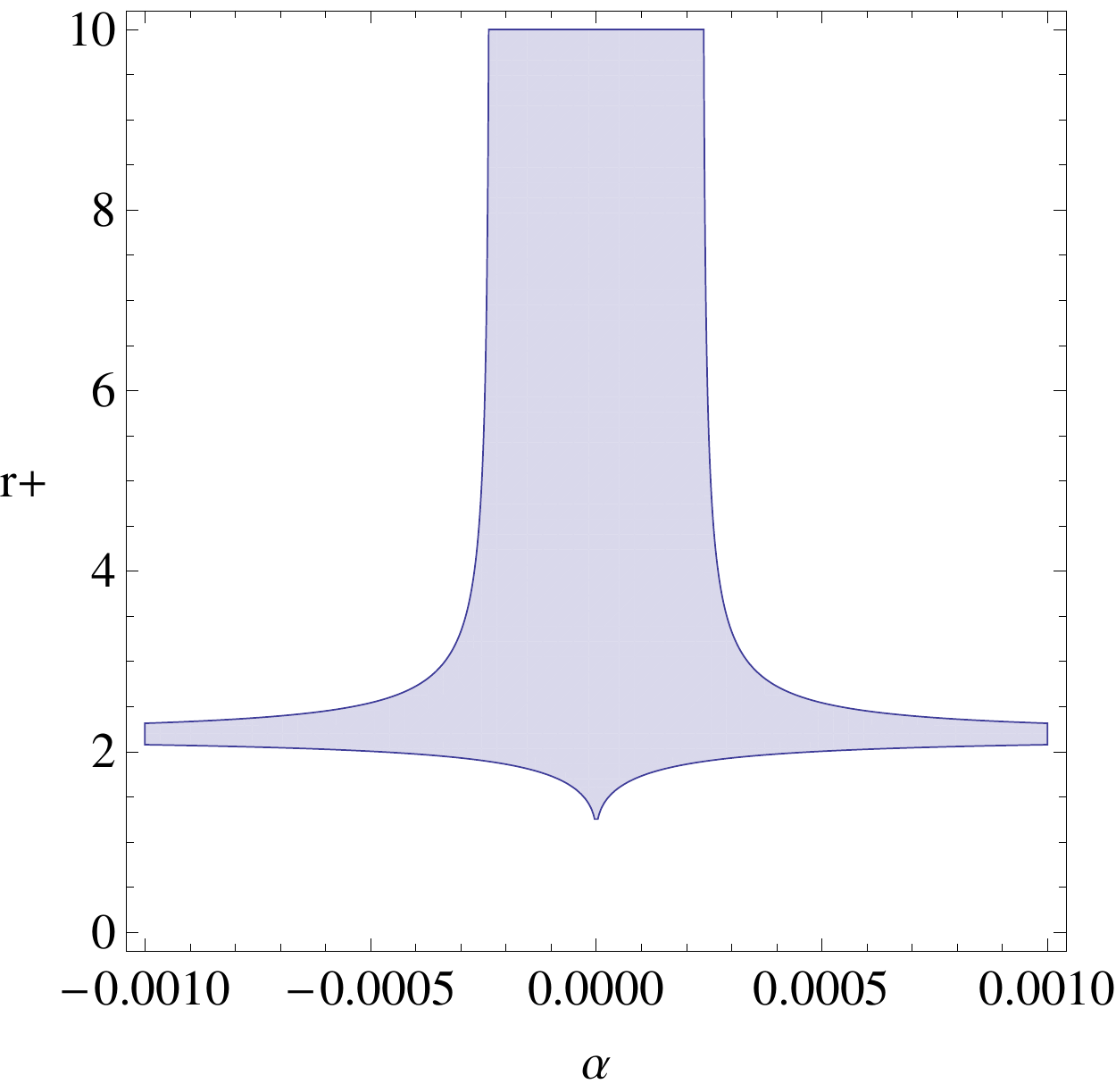}
\includegraphics[width=0.4\textwidth]{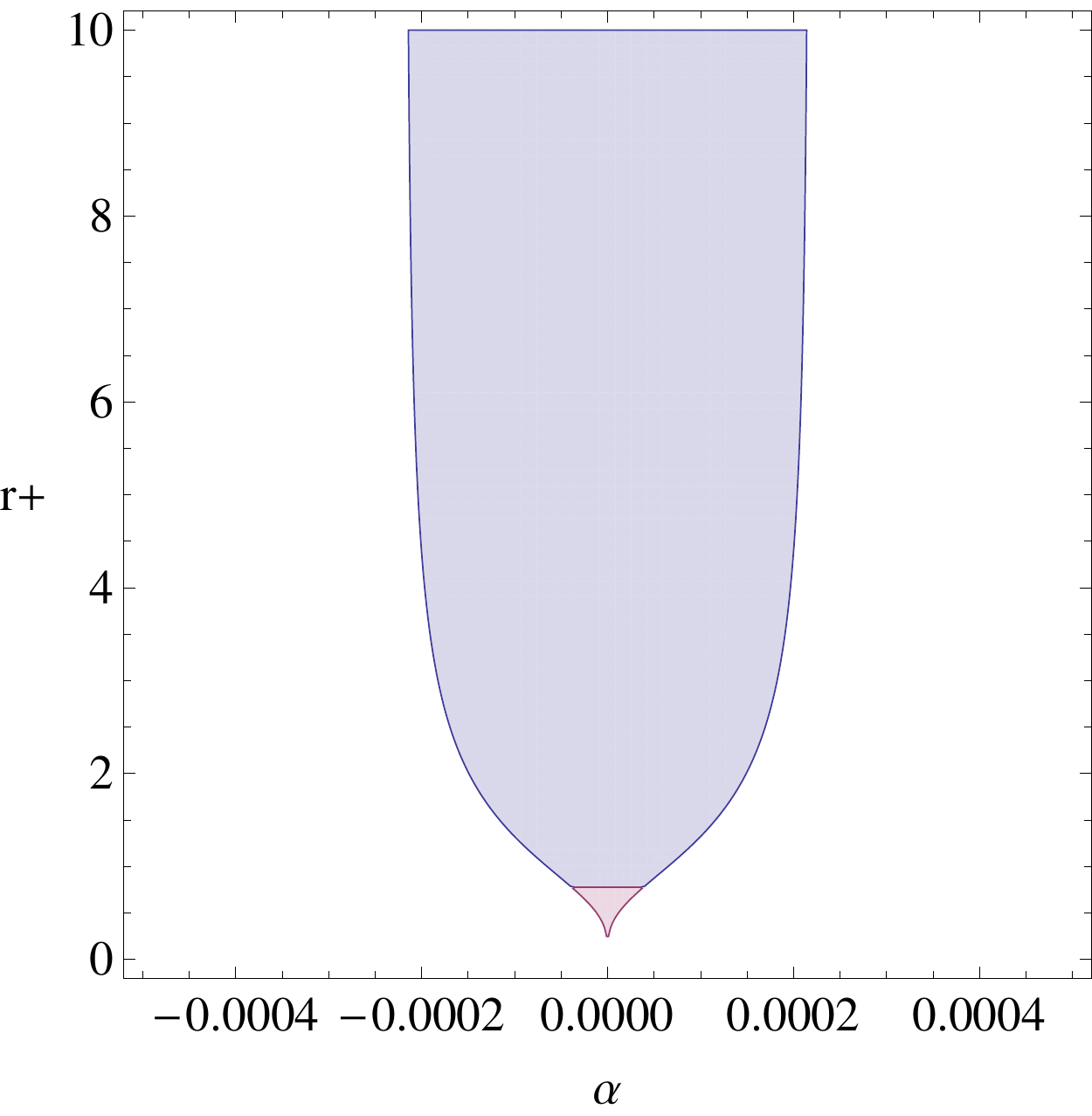}
\includegraphics[width=0.4\textwidth]{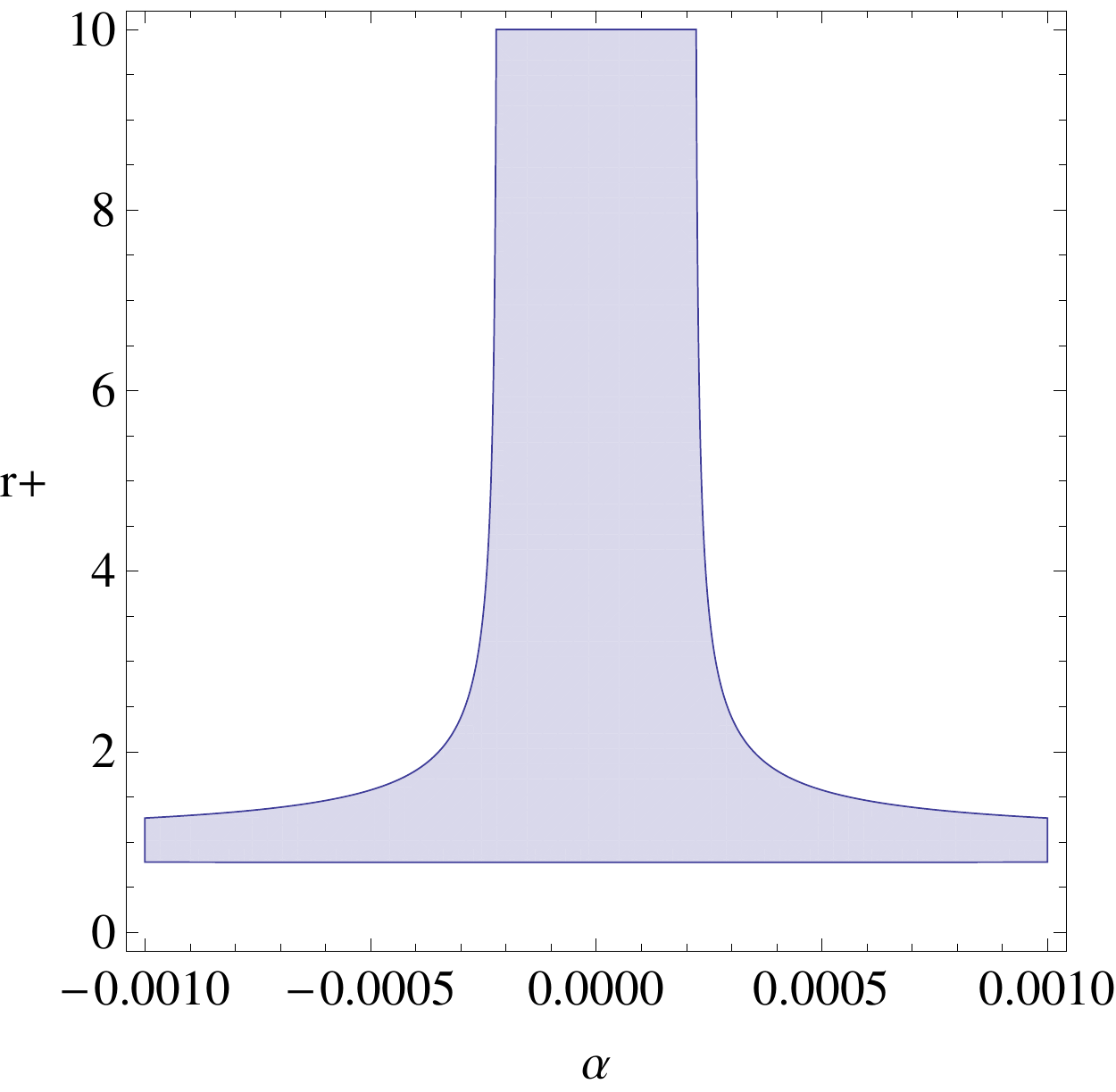}
\caption{{\bf Heat capacity}: Plots showing regions of parameter space where the heat capacity is and is not positive.  The plots are for $D=5$ with $k=+1$ and $k=-1$ (top left and right, respectively) and $D=6$ with $k=+1$ and $k=-1$ (bottom left and right, respectively).  In each case, the heat capacity is positive in the blue shaded regions and is negative in the red shaded regions.  In the production of these plots we have enforced $A_1 > 0$ and $\varepsilon |\alpha^2 A_1^{(2)}| > |\alpha^3 A_1^{(3)}|$ with $\varepsilon$ chosen as $1/10$.   }
\label{fig:heat_capacity}
\end{figure}

With an expansion for the mass it is also possible to examine the black hole solutions for thermodynamic stability.  Specifically, we can examine the specific heat 
\be 
C = \frac{\partial M}{\partial T}
\ee  
of the black holes and note when it is positive.  Examining this   to ${\cal O}(\alpha^2)$, we illustrate in  Figure~\ref{fig:heat_capacity}  the results of this investigation for $D=5$ and $D=6$ where the error in the series is less than 10\%.  In the regions where the heat capacity is negative (red shaded regions), these black holes are thermodynamically unstable.  We note that in the case of $k=-1$, the  black holes are thermodynamically stable with positive specific heat in the range where a second order expansion is valid.  

We close this section by studying the black hole equation of state for phase transitions in $D=5$ and $D=6$.  This can be obtained from Eq.~\eqref{eqn:nh_o1_higherd}, substituting for the temperature and pressure, i.e.~by returning to dimensionful units.  This is easily accomplished by defining
\be 
a_1 := \frac{A_1 r_+^2}{\ell^3}\, , \quad n_1 := \frac{N_1}{\ell} \, ,
\ee
after which Eq.~\eqref{eqn:nh_o1_higherd} reduces to the higher dimensional equivalent of Eq.~\eqref{a1_eq}.  Identifying the pressure as
\be 
P = -\frac{\Lambda}{\kappa} = \frac{(D-1)(D-2)}{2 \kappa \ell^2} \, ,
\ee
the equation of state then reads
\begin{align}\label{eqn:eos}
P =& \frac{2 \pi (D-2) T}{\kappa r_+} -\frac{ 96 \pi^2 \kappa (D-2)(D-3) k (4 r_+ n_1 + 1) \lambda T^2}{ r_+^4} 
\nn\\
& - \frac{64(D-2)(D-3)\left[ 12 r_+ n_1 + (D-4)  \right]\kappa \pi^3 \lambda T^3}{ r_+^3}
\nn\\
&+  \frac{(D-2)(D-3)k}{2\kappa r_+^6}\left[8 \kappa^2 k^2 (D-4)(D-5) \lambda -  r_+^4 \right]  
\end{align}
where $n_1$ is in general a free parameter, but is fixed for the Einstein branch as per Eq.~\eqref{eqn:some_first_order_expansions} in a way which is dependent on $P$.  Explicitly, to first order in $\lambda$, 
\be 
n_1 = \frac{6(D-1)(D-3)(D-4) \kappa^2 \lambda  \left(2 \kappa P r_+^2 + (D-1)(D-2)k \right)^2 }{(D-2)^2 r_+^5} \, .
\ee
Since this correction contains terms quadratic in $P$ it seriously complicates the study of the thermodynamics at order $\lambda^2$ and higher.  We can make progress by substituting for $P$ in the expression for $n_1$ above the full equation of state Eq.~\eqref{eqn:eos} and keep terms only to order $\lambda^2$.  We can then study the resulting effective equation of state for critical points, verifying whether or not the results are valid within the approximation.

Considering first the case of $D = 5$, we find that that there are no critical points when $k=+1$ working to second order in $\lambda$.  Naively, a calculation which is valid to first order suggests that there is a single critical point;  however, this critical point occurs for parameters at which the series approximation breaks down.  For $k=-1$ and $\lambda < 0$, there are two possible critical points occurring for the (dimensionless) combinations: $(\alpha, \rt, A_1) \approx (-0.05666, 0.91702, 5.608)$ and $(\alpha, \rt, A_1) \approx (-0.33412, 0.78780, 3.8581)$.  A quick calculation reveals that for the first case, while a series approximation is valid for the choice of $\alpha$ and $\rt$, the value of $A_1$ corresponds to a non-Einstein branch.  In the second case, the series approximation is not valid at the critical point.  With $k=-1$ and $\lambda > 0$, there is one possibility occurring at the dimensionless point $(\alpha, \rt, A_1) \approx (0.04595, 0.97383, 5.4786)$; however, once more the series approximation is not valid at this critical point.  

The situation is not much better for $D=6$.  For $k=+1$, working at ${\cal O}(\alpha^2)$ there are no options for critical points.  For $k=-1$, there are two possibilities corresponding to $(\alpha, \rt, A_1) \approx (0.59427, 0.55692, 0.36703)$ and $(\alpha, \rt, A_1) \approx (0.04387, 0.88529, .36703)$.  In the first case, the series approximation is not valid at the critical point. In the second case, while the series approximation is valid for the values of $\rt$ and $\alpha$, the value of $A_1$ indicates does not match that coming from the Einstein branch. For $k=-1$ and $\lambda < 0$, there is one possible critical point which occurs for $(\alpha, \rt, A_1) \approx (-0.05510, 0.90570, 6.3931)$ which does not correspond to the Einstein branch.  Thus, in five and six dimensions, there is no interesting critical behaviour that can be captured perturbatively.

\section{Conclusion}

 We have found new spherically symmetric vacuum black hole solutions of all topologies to Einsteinian cubic gravity using  a combination of analytic, perturbative, and numerical methods.  There is only ever a single ghost-free branch that  limits to the Einstein case when $\lambda \to 0$, and we find asymptotic AdS solutions in a broad region of parameter space.  We have found solutions in $D=4$ and
$D>4$.  The four dimensional solutions are described by a single function, $f(r)$, while the higher dimensional solutions require two independent functions.  

While these black holes have features and thermodynamic behaviour similar to their counterparts in Einstein and Lovelock gravity, they do exhibit some novel properties.  One is the existence of black holes of minimal radius, depending on the value coupling parameter $\alpha$.  Another is their equation of state, which is quadratic and cubic (and not linear) in the temperature.  In 4 dimensions their ratio of critical parameters is identical to that of a van der Waals fluid, the only such instance ever seen apart from the Reissner-Nordstrom black hole.  

 Turning to the higher dimensional cases, an interesting feature
 of the higher dimensional solutions is their violation (again in certain regions of parameter space) of the reverse isoperimetric inequality \eqref{RIPI}, the first time this has been observed for higher curvature black holes asymptotic to AdS space. And finally we note the existence of negative entropy solutions, whose physical status remains unclear, in significant regions of parameter space. 
 
 There remains a great deal of future work to be carried out for Einsteinian cubic gravity.  Within the theme of black hole thermodynamics, it would be interesting to see how the addition of the Gauss-Bonnet and cubic Lovelock terms affect the thermodynamics.  Furthermore, inclusion of matter sources, e.g. a Maxwell field, would add further structure to the thermodynamic behaviour, especially in higher dimensions.  More generally, it would be of interest to study holographic implications of the new cubic curvature term.  Of course, there remain more fundamental questions to be addressed as well, such as determining if the fourth-order nature of the field equations leads to any pathological behaviour beyond linear order.  

\section*{Acknowledgements}
This work was supported in part by the Natural Sciences and Engineering Research Council of Canada.  We are grateful to Pablo Bueno and Pablo Cano who pointed out to us an error in Section~\ref{sec:higherD} in an earlier version of this manuscript. 
\bigskip

{\bf Note added} In the course of preparing this manuscript we have discovered that a particular example of the five-dimensional hyperbolic solution presented here was recently studied in \cite{Dey:2016pei}.

\appendix

\section{${\cal O}(\alpha^2)$ corrections in higher dimensions}

Here we present the  ${\cal O}(\alpha^2)$ terms for the corrections to the metric functions for a selection of dimensions $D > 4$.  In particular, the metric function to  ${\cal O}(\alpha^2)$ reads,
\begin{align} 
f(\tilde r) &= 1 + \frac{k}{\tilde r^2} - \frac{c_0}{\tilde r^{d-1}} + \alpha h^{(D)}_1(\tilde r) + \alpha^2 h^{(D)}_2(\tilde r) \, ,
\nn\\
N(\tilde r) &= 1 + \alpha N_1^{(D)}(\tilde r) + \alpha^2 N_2^{(D)}(\tilde r) \, ,   
\end{align}
where $h^{(D)}_1(\tilde r)$ and $N^{(D)}_1(\tilde r)$ are given by Eq.~\eqref{fo_correction} and $ h^{(D)}_2(\tilde r)$ and $N^{(D)}_2(\tilde r)$ in 5, 6 and 7 dimensions are given by:
\allowdisplaybreaks
\begin{align}
h^{(D=5)}_2(\tilde r)=& {\frac {16}{27}}+{\frac {c_{{2}}}{{\tilde r}^{4}}}-{\frac {1472\,c_{{0}}^{2
}}{27\,{\tilde r}^{8}}}-{\frac {184\,c_{{0}}c_{{1}}}{3\,{\tilde r}^{8}}}-{\frac {256
\,k c_{{0}}^{2}}{9\,{\tilde r}^{10}}}-{\frac {128\,kc_{{0}}c_{{1}}}{3\,{\tilde r}^{
10}}}-{\frac {167696\,c_{{0}}^{3}}{27\,{\tilde r}^{12}}}+{\frac {148\,c_{{1
}}c_{{0}}^{2}}{3\,{\tilde r}^{12}}}
\nn\\
&-11776\,{\frac {kc_{{0}}^{3}}{{\tilde r}^{14}
}}-{\frac {51200\,{k}^{2}c_{{0}}^{3}}{9\,{\tilde r}^{16}}}+12080\,{\frac {
c_{{0}}^{4}}{{\tilde r}^{16}}}+{\frac {81664\,k c_{{0}}^{4}}{7\,{\tilde r}^{18}}}-
{\frac {160064\,c_{{0}}^{5}}{27\,{\tilde r}^{20}}}
\\
N^{(D=5)}_2(\tilde r)= & -\,{\frac {8c_{{0}}^{2}n_{{1}}}{3 {\tilde r}^{8}}}+{\frac {32\,c_{{0
}}^{2}}{9\,{\tilde r}^{8}}}+\,{\frac {16 c_{{0}}c_{{1}}}{3{\tilde r}^{8}}}+{\frac {
17792\,c_{{0}}^{3}}{27\,{\tilde r}^{12}}}+{\frac {51200\,kc_{{0}}^{3}}{63
\,{\tilde r}^{14}}}-{\frac {9200\,c_{{0}}^{4}}{9\,{\tilde r}^{16}}} +  n_{{2}}
\\
h^{(D=6)}_2(\tilde r)=& {\frac {264222\,kc_{{0}}^{4}}{17\,{\tilde r}^{22}}}+{\frac {117\,c_{{1}}c_
{{0}}^{2}}{2\,{\tilde r}^{15}}}-15552\,{\frac {kc_{{0}}^{3}}{{\tilde r}^{17}}}-{
\frac {37908\,{k}^{2}c_{{0}}^{3}}{5\,{\tilde r}^{19}}}-72\,{\frac {c_{{0}}c
_{{1}}}{{\tilde r}^{10}}}-54\,{\frac {kc_{{0}}c_{{1}}}{{\tilde r}^{12}}}
\nn\\
&+15903\,{
\frac {c_{{0}}^{4}}{{\tilde r}^{20}}}-{\frac {63207\,c_{{0}}^{5}}{8\,{\tilde r}^
{25}}}-8100\,{\frac {c_{{0}}^{3}}{{\tilde r}^{15}}}+{\frac {c_{{2}}}{{\tilde r}^{5
}}}
\\
N^{(D=6)}_2(\tilde r)= &-\,{\frac {9c_{{0}}^{2}n_{{1}}}{2 {\tilde r}^{10}}}+9\,{\frac {c_{{0
}}c_{{1}}}{{\tilde r}^{10}}}+1242\,{\frac {c_{{0}}^{3}}{{\tilde r}^{15}}}+{\frac {
123444\,kc_{{0}}^{3}}{85\,{\tilde r}^{17}}}-{\frac {14985\,c_{{0}}^{4}}{8
\,{\tilde r}^{20}}} +  n_{{2}}
\\
h^{(D=7)}_2(\tilde r)=& {\frac {1024}{16875}}+{\frac {c_{{2}}}{{\tilde r}^{6}}}+{\frac {372736\,c_{{0
}}^{2}}{16875\,{\tilde r}^{12}}}-{\frac {5824\,c_{{0}}c_{{1}}}{75\,{\tilde r}^{12}}
}+{\frac {8192\,kc_{{0}}^{2}}{625\,{\tilde r}^{14}}}-{\frac {1536\,kc_{{0}}
c_{{1}}}{25\,{\tilde r}^{14}}}
\nn\\
&-{\frac {155762176\,c_{{0}}^{3}}{16875\,{\tilde r}^{
18}}}+{\frac {4784\,c_{{1}}c_{{0}}^{2}}{75\,{\tilde r}^{18}}}-{\frac {
11206656\,kc_{{0}}^{3}}{625\,{\tilde r}^{20}}}-{\frac {5505024\,{k}^{2}c_{
{0}}^{3}}{625\,{\tilde r}^{22}}}
\nn\\
&+{\frac {309192704\,c_{{0}}^{4}}{16875\,{\tilde r
}^{24}}}+{\frac {56426496\,kc_{{0}}^{4}}{3125\,{\tilde r}^{26}}}-{\frac {
155042944\,c_{{0}}^{5}}{16875\,{\tilde r}^{30}}}
\\
N^{(D=7)}_2(\tilde r)= & -{\frac {144\,c_{{0}}^{2}n_{{1}}}{25\,{\tilde r}^{12}}}-{\frac {1536
\,c_{{0}}^{2}}{625\,{\tilde r}^{12}}}+{\frac {288\,c_{{0}}c_{{1}}}{25\,{\tilde r}^
{12}}}+{\frac {1042432\,c_{{0}}^{3}}{625\,{\tilde r}^{18}}}+{\frac {5935104
\,kc_{{0}}^{3}}{3125\,{\tilde r}^{20}}}
\nn\\
&-{\frac {311424\,c_{{0}}^{4}}{125
\,{\tilde r}^{24}}} + n_{{2}}
\end{align}
\bibliography{LBIB}
\bibliographystyle{JHEP}

\end{document}